\documentclass[preprint,apj]{emulateapj}

\newcommand{\sersic}{S\'ersic}
\newcommand{\deVa}{de Vaucouleurs}
\newcommand{\HST}{{\em HST}}
\newcommand{\ACS}{{\em ACS}}

\newcommand{\gems}{{\sc Gems}}
\newcommand{\stages}{{\sc Stages}}
\newcommand{\cosmos}{{\sc Cosmos}}

\newcommand{\goods}{{\sc Goods}}
\newcommand{\galapagos}{{\sc Galapagos}}
\newcommand{\galfit}{{\sc Galfit}}
\newcommand{\gasphot}{{\sc Gasphot}}
\newcommand{\magarc}{{magnitudes\ arcsec$^{-2}$}}
\newcommand{\gimtwod}{{\sc Gim2d}}
\newcommand{\sex}{{\sc SExtractor}}

\usepackage{lscape}

\shorttitle{\gems: Galaxy fitting catalogues and testing parametric galaxy fitting
codes} \shortauthors{H\"au\ss ler et al.}

\received{2006 Sept 12}
\begin{document}

\title{\gems: Galaxy fitting catalogues and testing
parametric galaxy fitting codes:\\
       \galfit, \gimtwod}

\author{
Boris H\"au\ss ler\altaffilmark{1},
Daniel H.~McIntosh\altaffilmark{2},
Marco Barden\altaffilmark{1},
Eric F.~Bell\altaffilmark{1},
Hans-Walter Rix\altaffilmark{1},
Andrea Borch\altaffilmark{3},
Steven V.~W.~Beckwith\altaffilmark{4,5},
John A.~R.~Caldwell\altaffilmark{6},
Catherine Heymans\altaffilmark{7},
Knud Jahnke\altaffilmark{1},
Shardha Jogee\altaffilmark{8},
Sergey E.~Koposov\altaffilmark{1},
Klaus Meisenheimer\altaffilmark{1},
Sebastian F.~S\'anchez\altaffilmark{9},
Rachel S.~Somerville\altaffilmark{1},
Lutz Wisotzki\altaffilmark{10},
Christian Wolf\altaffilmark{11} }
\email{boris@mpia.de}

\altaffiltext{1}{Max-Planck-Institut f\"ur Astronomie,
K\"onigstuhl 17, 69117, Heidelberg, Germany}
\altaffiltext{2}{Department of Astronomy,
University of Massachusetts, 710 North Pleasant Street, Amherst, MA
01003, USA}
\altaffiltext{3}{Astronomisches Recheninstitut,
M\"onchhofstra\ss e 12-14, 69120, Heidelberg, Germany}
\altaffiltext{4}{Space Telescope Science Institute, 3700 San Martin Dr.,
Baltimore, MD 21218, USA}
\altaffiltext{5}{Johns Hopkins University, 3400
North Charles Street, Baltimore, MD 21218, USA}
\altaffiltext{6}{University of Texas, McDonald Observatory, Fort Davis,
TX 79734, USA}
\altaffiltext{7}{Department of Physics and Astronomy,
The University of British Columbia, 6224 Agricultural Road Vancouver, V6T
1Z1 Canada}
\altaffiltext{8}{University of Texas at Austin, 1, University
Station C1400, Austin, TX 78712-0259, USA}
\altaffiltext{9}{Centro
Astronomico Hispano Aleman de Calar Alto, C/Jesus Durban Remon 2-2,
Almeria, E-04004, Spain}
\altaffiltext{10}{Universit\"at Potsdam, Am
Neuen Palais 10, 14469, Potsdam, Germany}
\altaffiltext{11}{Department of
Physics, Denys Wilkinson Bldg., University of Oxford, Keble Road, Oxford,
OX1 3RH, UK}

\begin{abstract}
In the context of measuring structure and morphology of intermediate redshift
galaxies with recent {\it HST}/\ACS\ surveys, we tune, test, and compare two widely
used fitting codes (\galfit\ and \gimtwod) for fitting single-component \sersic\
models to the light profiles of both simulated and real galaxy data. Our study
focuses on the \gems\ survey with the sensitivity of typical {\it HST} survey data,
and we include our final catalog of fit results for all 41,495 objects detected in
\gems. We test the reliability of both codes using simulated galaxy profiles
constructed to represent the imaging characteristics of \gems. We find that fitting
accuracy depends sensitively on galaxy profile shape. Exponential disks are well fit
with \sersic\ models and have small measurement errors, whereas fits to \deVa\
profiles show larger uncertainties owing to the large amount of light at large
radii. We find that both codes provide reliable fits and little systematic error,
when the effective surface brightness is above that of the sky. Moreover, both codes
return errors that significantly underestimate the true fitting uncertainties, which
are best estimated with simulations. We find that \gimtwod\ suffers significant
systematic errors for spheroids with close companions owing to the difficulty of
effectively masking out neighboring galaxy light; there appears to be no work around
to this important systematic in \gimtwod's current implementation. While this
crowding error affects only a small fraction of galaxies in \gems, it must be
accounted for in the analysis of deeper cosmological images or of more crowded
fields with \gimtwod. In contrast, \galfit\ results are robust to the presence of
neighbors because it can simultaneously fit the profiles of multiple companions
thereby deblending their effect on the fit to the galaxy of interest. We find
\galfit's robustness to nearby companions and factor of $\gtrsim20$ faster runtime
speed are important advantages over GIM2D for analyzing large {\it HST}/\ACS\
datasets.
\end{abstract}


\setcounter{footnote}{0}
\section{Introduction}\label{sec_Introduction}

One of the central goals of observational exploration of galaxy evolution is to
understand how the structures of galaxies evolve with cosmic time. A powerful tool
in this context are large look-back surveys, where the time evolution of the
distribution of galaxy structural properties can be quantified. The key to unlocking
the potential of these surveys is the development of quantifiable, well-understood,
and repeatable ways to measure and describe galaxy structures. Using such methods,
the evolution of the structure of disk galaxies
\citep{Lilly1998,Simard1998.2,Ravindranath,Barden2005,Trujillo2005,Sargent} and
spheroid-dominated galaxies
\citep[e.g.,][]{Schade97,Schade99,McIntosh2005,Trujillo2004,Trujillo2006} has been
quantified over the last 10 billion years of cosmic time, since $z=3$. In this
paper, we exhaustively test and tune two parametric galaxy fitting codes, \galfit\
\citep{Peng} and \gimtwod\ \citep{Simard02}, that are commonly used in the
literature. With these tests we determine the best fitting setups for each code,
quantify the sources of random and systematic uncertainty, and presents parametric
fits for 41,495 objects in the \HST\ \gems\ \citep{Rix} dataset.

There are two main approaches towards describing galaxy structure from the
two-dimensional information contained in image data. Non-parametric methods provide
estimates of total brightness, galaxy half-light size, and structure, using metrics
which do not depend on a galaxy, having a structure well-described by any particular
functional form \citep[e.g.,][]{Petrosian,Abraham,Bershady,Lotz}. The main
disadvantages of non-parametric methods are that they are reasonably sensitive to
the depth of the images; because there is no parametric form for extrapolating to
account for the faint outer parts of galaxies, one can underestimate flux and/or
size in poorly-posed cases \citep{Blanton}. Parametric methods, in contrast, choose
particular functional forms (sometimes reasonably complicated) with which to fit the
galaxy light distribution. These have substantially less flexibility than
non-parametric fitting codes, but have the advantage that light at large radii can
be accounted for reasonably well by the natural extrapolation of the best-fitting
model profile (under the assumption that the parametric form chosen does, in fact,
describe the light profile in the outer parts of galaxies reasonably well). Besides
robust estimates of galaxy size, parametric methods provide measures of galaxy
structure that may shed light on relative contributions of physically distinct and
meaningful components such as spheroids, disks, and stellar bars.

\begin{figure}[htb]
\begin{center}
\includegraphics[width=8cm,angle=0]{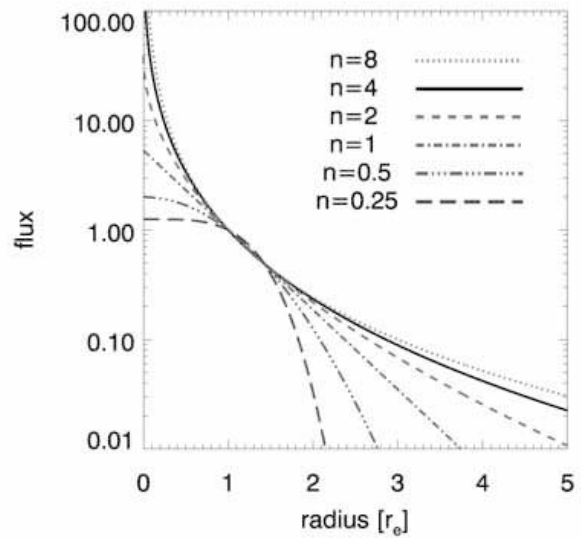}
\end{center}
\caption{This plot shows \sersic\ profiles for different values of the \sersic\
index $n$, normalized to have the same flux at $r_{e}$. One can see that profiles
with high \sersic\ indices $n\gtrsim2$ have more flux at larger radii; thus, a good
estimate of the background sky level is particularly important for precise fitting
of high-$n$ galaxies. \label{sersic_profiles}}
\end{figure}

One particularly useful and flexible profile for parametric galaxy
fitting is a single-component \citet{Sersic} model, which describes the
radial surface brightness profile of a galaxy by the \sersic\ function
given by
\begin{equation}
\Sigma(r)=\Sigma_{e} \cdot exp\ [-\kappa((r/r_{e})^{1/n}-1)]  ,
\label{eqn:ser}
\end{equation}
where $r_{e}$ is the radius of the galaxy (Note that for a \sersic\ fit $r_{\rm e}$
is equivalent to the half-light radius $r_{50}$), $\Sigma_{e}$ is the surface
brightness at $r_{e}$, and the \sersic\ parameter $n$ describes the profile shape
(the parameter $\kappa$ is closely connected to $n$).  Together with position (x and
y), axis ratio $b/a$ and position angle, this profile has 7 free parameters. The
\sersic\ profile represents a more general form of the exponential light-profiles
seen in galactic disks ($n=1$) and the $R^{1/4}$-law (\deVa\ law) profiles typical
of luminous early-type galaxy ($n=4$) \citep[e.g.,][]{deVac, Freeman}; fitting with
this profile has been explored in detail in a number of works
\citep[e.g.,][]{Simard98,Simard02,Graham,Trujillo2001}. Figure \ref{sersic_profiles}
shows some examples of \sersic\ profiles with different $n$. Many authors have used
a constant value of $n=2.5$ or $n=2.0$ to crudely distinguish early-type
(bulge-dominated) from late-type (disk-dominated) galaxies
\citep[e.g.,][]{Blanton2003,shen03,hogg04,Bell2004,Barden2005,McIntosh2005,
Ravindranath}. Furthermore, fitting galaxies with a \sersic\ profile gives an
estimate of size, and therefore is very useful for the examination of the evolution
of galaxy scaling relations.

The goal of this paper is to describe our efforts to optimize the estimation of
single-component \sersic\ profile fits to the galaxies in the \gems\ survey
\citep{Rix}. To date, this has been our primary method for quantifying galaxy
structure\footnote{Bulge-disk composite galaxies were {\it not} simulated for this
paper; bulge-disk decomposition will be addressed in a future paper.}. We compare
the performance of the \galfit\ and \gimtwod\ automated galaxy fitting codes, which
are suitable for fitting large datasets such as \gems, \stages\ \citep[][in
prep.]{gray06}, \goods\ \citep{Giavalisco} and \cosmos\ \citep{Scoville}. We address
the reliability and limitations of these codes through thorough testing, using
simulated and real galaxies. We describe the details of the simulations used
throughout this paper in \S\ref{sec_sims}. In \S\ref{sec_codes} we explore different
set-ups, converging on `best-fitting' set-ups for each fitting code.
\S\ref{sec_results} summarizes the results from our testing of these `best-fitting'
set-ups using both simulated galaxies (\S\ref{sec_results_sims} and
\S\ref{sec_results_fake}) and real galaxies (\S\ref{sec_results_deep}). We compare
our findings with those of a recent paper on the same topic (using the codes
\galfit, \gimtwod, and \gasphot) by \citet{Pignatelli} in \S\ref{sec_gasphot}, and
publish a catalog of \galfit\ fitting results for all 41,495 detected galaxies from
the \gems\ survey in \S\ref{sec_gems_results}. All results from this paper and a
number of other catalogs and images useful for testing galaxy fitting codes are
presented on the \gems\ webpage\footnote{see http://www.mpia.de/GEMS/gems.htm}.

\section{Simulations}\label{sec_sims}

Galaxy simulations are an invaluable tool for understanding the performance of
quantitative fitting pipelines. In this section we describe the set of simulations
that are extensively used for this paper; the results obtained from fitting these
simulations are discussed in \S\ref{sec_results}.

In this paper, we focus on simulations of two different galaxy light profiles:
purely exponential profiles ($n=1$) representing the luminosity profile of a
galactic disk (we will call these galaxies `disks' throughout this paper), and
\sersic\ profiles with a \sersic\ index of $n=4$ representing a \deVa\ luminosity
profile of a galactic bulge/elliptical galaxy (we will call them `spheroids'),
respectively. Profiles having \sersic\ indices between these two values of 1 and 4
are not presented here because $n=1$ and $n=4$ simulations span the range of
observed behavior, exponential profiles being the `easiest' to fit, \deVa\ being the
`hardest'. Nonetheless, extensive simulations of intermediate profiles (200,000
objects) have been produced, the catalog of \galfit\ fitting results for this sample
can also be downloaded from the \gems\ webpage.

This section is arranged as follows. The simulation of individual galaxies is
described in \S\ref{noisefree}. Section \ref{sim_gems} describes the construction of
simulated \gems\ frames from the individual galaxy simulations, including the
addition of realistic noise.

\subsection {Simulation of individual noise-free galaxies \&
Oversampling} \label{noisefree}

Galaxies were simulated using a custom-built IDL routine. Most available standard
routines (like \verb+MKOBJECT+ in \verb+IRAF+, \verb+create/image+ in \verb+MIDAS+
and similar tasks in other programs) compute the correct flux value for the {\it
center} of the pixel, but due to curvature of the profile, taking this as the mean
flux value for the {\it whole} pixel is incorrect. The higher the curvature is
(within a certain pixel), the more one underestimates the true pixel value. This
implies progressively larger inaccuracies for higher \sersic\ indices.

While it is possible to analytically integrate the profile across a pixel to obtain
an exact answer, this procedure is very CPU-intensive. We adopt a hybrid approach.
We use IDL's \verb+dist_ellipse+ routine in conjunction with equation
(\ref{eqn:ser}) to compute \sersic\ galaxy models which, as in the above cases, are
only correct for the center of the pixel. In order to increase accuracy, the inner
parts of our simulated profiles (100x100 pixels up to 200x200 pixels depending on
object size) have been oversampled by a factor of 10, and the very inner parts
(10x10 up to 20x20 pixels) are oversampled by a factor of 100. This was done by
creating the images by a factor of 10 (or 100 respectively) bigger and then
rebinning the image while holding the total flux constant. In this way, it is
possible to create a final profile accurate to better than 0.03\% at all radii (much
smaller than the poisson noise added later in the simulation process) with a factor
of 100 gain in speed compared to the analytical integration -- an important gain
when simulating large samples of galaxies.

\subsection {Simulation of crowded images} \label{sim_gems} To
realistically test galaxy extraction and fitting codes requires the creation of
images with large numbers of simulated galaxies distributed as in real data. Such
images were created by providing a catalog of simulated galaxy input parameters to
the simulation code, which simulated galaxies at the location, luminosity, size,
orientation and axis ratio $b/a$ specified in this catalog. In this step, galaxies
were put in an empty image of the same size as the final image.

\begin{deluxetable*}{lccl}
\tablewidth{0pt} \tablenum{1} \tabletypesize{\scriptsize}
\tablecolumns{4} \tablecaption{Simulation parameters, disk ($n$=1)
galaxies} \tablehead{\colhead{Parameter} & \colhead{Min} & \colhead{Max}
& \colhead{Distribution}} \startdata
mag [mag]         & 20 &  26.5 &  uniform\\
$r_{e}$ [pixel]   & 2 & 316 &  uniform in logarithmic space\\
   & & & $r_{e}< 10^{7.36 - 0.233\cdot mag}$, mag being chosen magnitude for object \\
$b/a$                & 0.18 &  1 &  uniform in $\cos(i)$, $i$ being inclination angle\\
   & & &  corrected for intrinsic thickness: \\
   & & &  $b/a=\sqrt{\cos^{2}(i)+(\sin(i)\cdot 0.18)^2}$\\
   & & &  intrinsic thickness ~0.18 following \citet{Pizagno, Ryden} and others\\
PA [deg to image] & 0 & 180 &  uniform\\
 \sersic\ index $n$    & 1 &  1 & fixed\\
\enddata
\label{tab_sim_param_disk}
\end{deluxetable*}
\begin{deluxetable*}{lccl}
\tablewidth{0pt} \tablenum{2} \tabletypesize{\scriptsize}
\tablecolumns{4} \tablecaption{Simulation parameters, spheroidal ($n$=4)
galaxies} \tablehead{\colhead{Parameter} & \colhead{Min} & \colhead{Max}
& \colhead{Distribution}} \startdata
mag [mag]         & 20 &  27 &  uniform\\
$r_{e}$ [pixel]   & 2 & 630 & uniform in logarithmic space\\
   & & & $r_{e}< 10^{4.79 - 0.1\cdot mag}$ \\
   & & & $r_{e}< 10^{11.49 - 0.392\cdot mag}$ \\
$b/a$                & 0.45 &  1 &  uniform in $\cos(i)$, $i$ being inclination angle\\
   & & &  corrected for intrinsic thickness: \\
   & & &  $b/a=\sqrt{\cos^{2}(i)+(\sin(i)\cdot 0.45)^2}$\\
PA [deg to image] & 0 & 180 &  uniform\\
 \sersic\ index $n$    & 4 &  4 & fixed\\
\enddata
\label{tab_sim_param_bulge}
\end{deluxetable*}
To choose the range of galaxy parameters for the simulated galaxies, we first fitted
all \gems\ galaxies with \galfit\ and determined the parameter range covered by the
real galaxy sample. Given these results, we chose a wider range of parameter space
for the simulations, in order to test detection efficiency, completeness, and to
allow pushing both parametric fitting codes to their limits. The simulations have a
random distribution in size between 2 and 316 pixels (uniformly distributed in
logarithmic space) and magnitude between 20 and 27 (uniform). With this distribution
of parameters, there were a relatively large number of large and low surface
brightness galaxies (stringently testing the detection efficiency and fitting
codes); we discuss this point in more detail in \S\ref{sec_results_fake}. The exact
distributions of simulation parameters are given in Tables \ref{tab_sim_param_disk}
and \ref{tab_sim_param_bulge}.

After simulating the galaxy profiles and putting them in an empty image, this final
image was convolved with a real F850LP-band PSF derived from the \gems\ dataset
\citep{Jahnke}. Next, an appropriate amount of noise had to be added to the images.
Owing to the multiple-frame dither characteristic of \HST\ imaging surveys, the
noise is somewhat correlated pixel-to-pixel. Thus, strictly speaking, galaxies
should be simulated in individual dithers, then dithered together using exactly the
same routines as were used to combine the \gems\ frames. We took an intermediate
approach: Poisson noise with the same RMS as the \gems\ noise was added to the
simulated galaxy frame, then a real `sky' frame was added to the simulated frame to
accurately account for real fluctuations and correlated noise in observed \HST\ sky
backgrounds. We have confirmed through tests with \galfit\ that this (much less CPU-
and work-intensive) hybrid approach yields a scatter which is negligible compared to
random fitting uncertainties. The `sky' frame was constructed by adding \gems\
F606W- and F850LP-band frames (to increase image depth and to make sure that objects
appear in neither of the two bands) and visually checking those images to identify
patches of 500x500 pixels in size without objects detected by \sex. The chosen
patches were cut from the F850LP-band images and pasted together to form an empty
image of the same size as an original \gems\ image.

\begin{figure*}[htb]
\begin{center}
\includegraphics[width=8cm,angle=0]{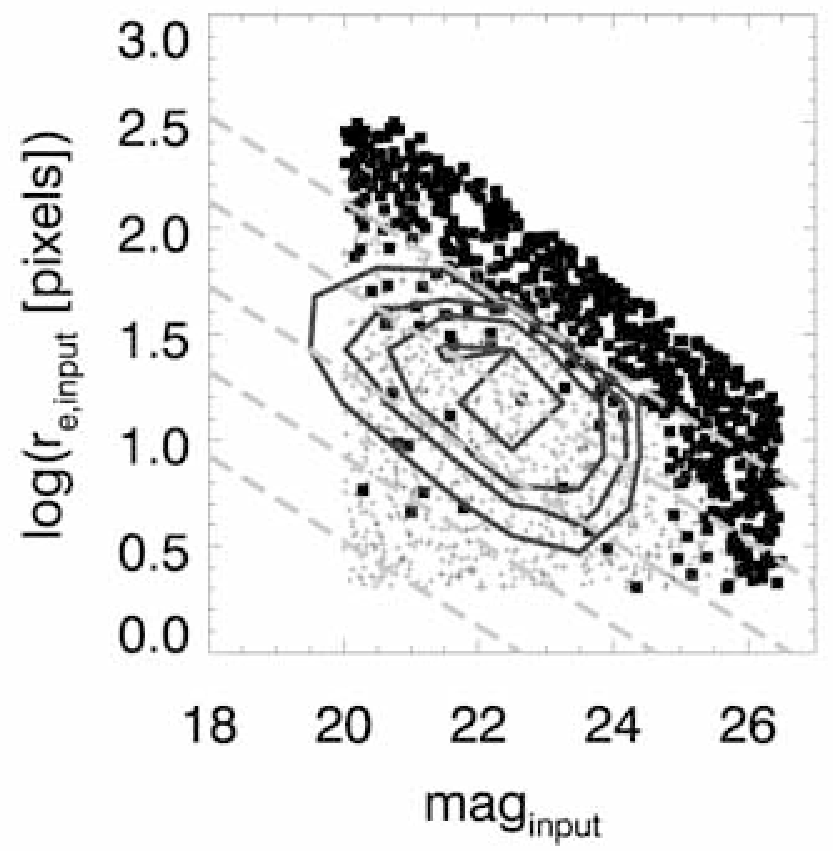}
\includegraphics[width=8cm,angle=0]{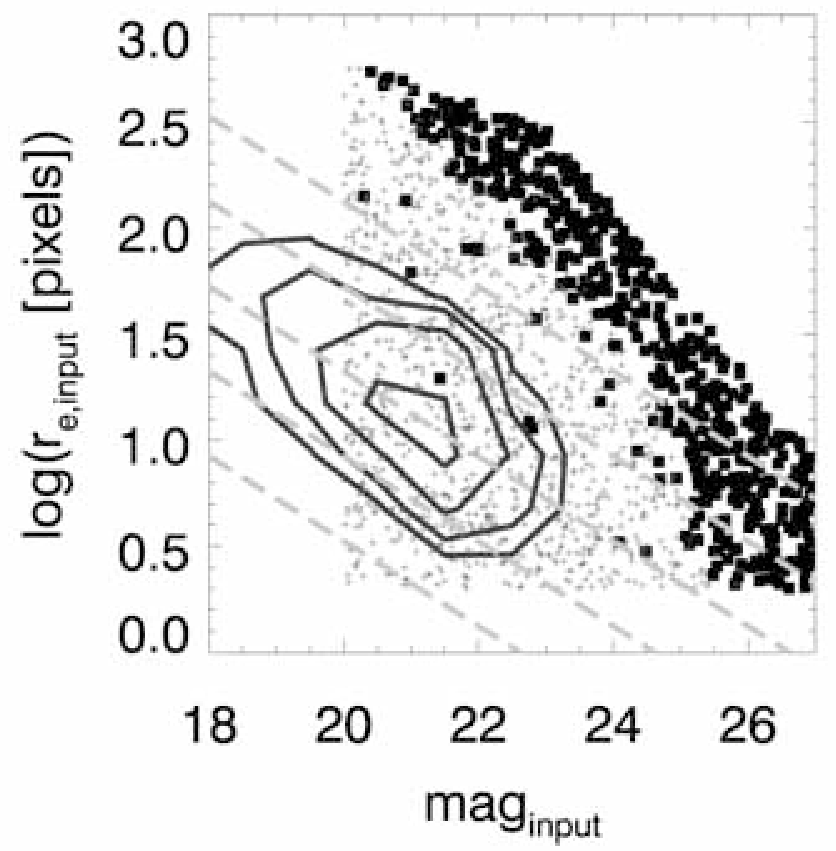}
\end{center}
\caption{F850LP magnitudes and sizes for the full set (all symbols) of 1600
simulated $n$=1 galaxies (left) and $n$=4 galaxies (right), and the subsets that
were detected by \sex\ (small grey crosses). The black squares indicate galaxies
that were missed during object detection. The indicated contours show the
magnitude-size space populated by actual \gems\ galaxies used by \citet[for disk
galaxies][left]{Barden2005} and \citet[for spheroidal
galaxies][right]{McIntosh2005}; the contours show the areas of parameter space where
the reliability of the fitting routines becomes especially important. Whereas real
$n\geq2.5$ galaxies lie in the area where all galaxies are detected, we did use
$n\leq2.5$ galaxies that are close to the edge of detectability for our analysis.
The different behavior of the non-detected galaxies in both samples reflects the
fact that, due to their bright central peak, galaxies with a high \sersic\ index are
easier to detect than galaxies with low $n$. To guide the eye, we overplot
long-dashed lines of constant surface brightness of 17, 19, 21, 23 and 25 \magarc\
from bottom to top. \label{fig_completeness}}
\end{figure*}

The result of the simulation process was a simulated image with noise properties
very similar to a real \gems\ image that contained 800 simulated light profiles with
different magnitudes, sizes, position angles and values of $b/a$. Different sets of
simulations were created in this way: one set contains disk-like $n=1$ galaxies only
(for results see \S\ref{sec_results_sims_disk}) and the other contains spheroidal
{\it $n=4$} profiles only (see \S\ref{sec_results_sims_bulge}). Not all 800 galaxies
were recovered by \sex. Roughly 80\% of the objects were recovered, depending on the
simulated profile shape and the distribution of galaxy parameters in the particular
image (see Figure \ref{fig_completeness}). In particular, very large and low
surface-brightness galaxies were not detected \citep[see Figure
\ref{fig_completeness} and ][]{Rix}. Due to the fact that spheroids are easier to
detect due to their centrally concentrated light profiles, less galaxies were
recovered in the disk sample.

\begin{figure*}[htb]
\begin{center}
\includegraphics[width=8cm,angle=0]{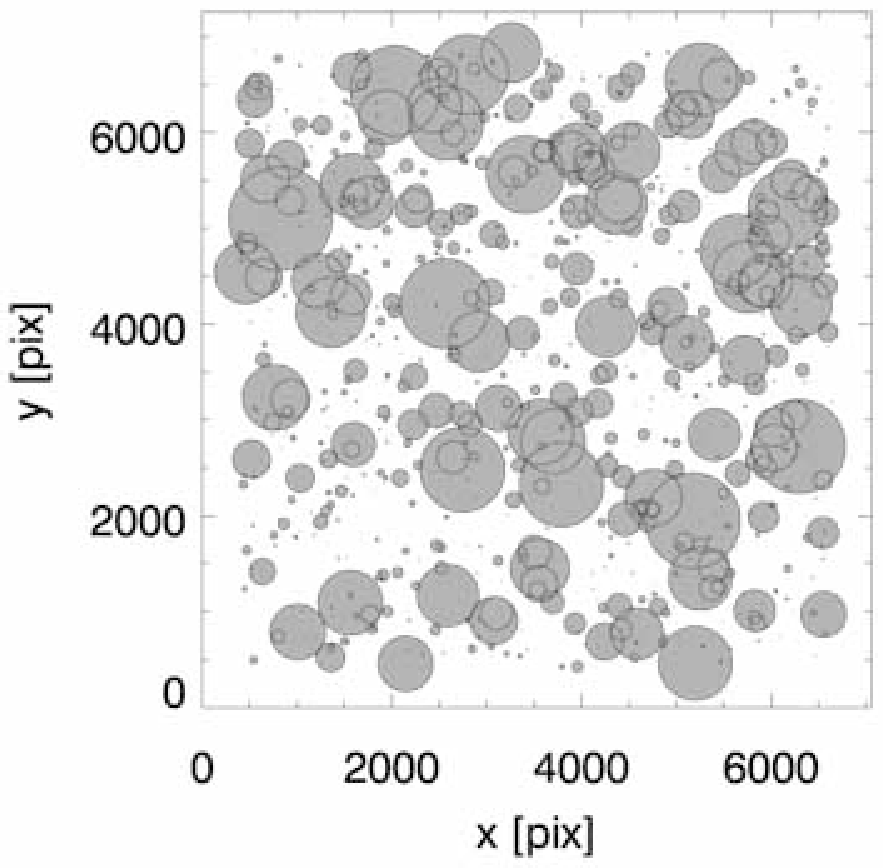}
\includegraphics[width=8cm,angle=0]{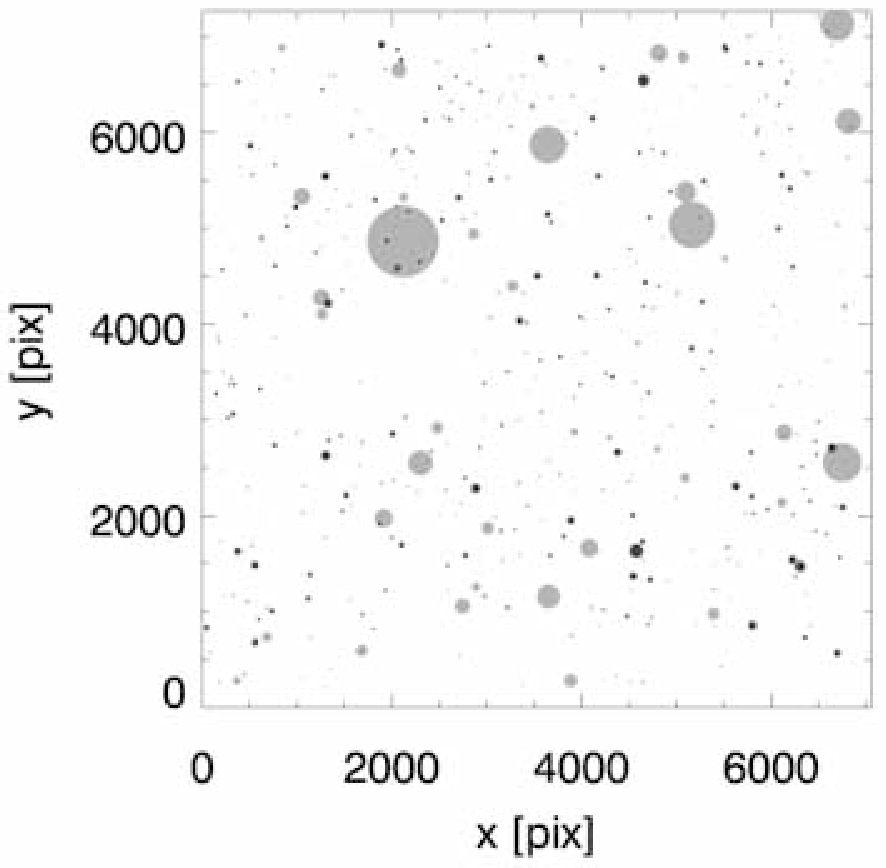}
\end{center}
\caption{Density of sources in simulation images. {\it Left:} One of the two
simulated images with galaxies having \sersic\ index of 4. We plot a circle with
radius $r_{e}$ (simulated) at the correct $x,y$ location of each simulated galaxy.
In total there are 800 simulated $n=4$ galaxies placed in a single \ACS\ tile,
$\sim250$ of which are too LSB to be detected by \sex; they could nonetheless
influence the fitting results by contributing to the image background. This
simulation represents an extreme case for testing the limits of profile fitting with
\gimtwod\ and \galfit. {\it Right:} the sources in a simulation of a typical \gems\
\ACS\ image (\gems\ tile 04) using \galfit\ fitting results. There are 523 total
simulated galaxies, 374 with $n$=1 (in dark grey) and 148 with $n$=4 (light grey).
Stars and the few objects (in total 46 objects) that ran into a fitting constraint
where excluded from this simulation. One can easily see from this plot that real
galaxies are significantly less crowded than the completely artificial simulations
used in \S\ref{sec_results_sims_disk} and \S\ref{sec_results_sims_bulge}
\label{fig_crowding}}
\end{figure*}

By design, these simulated tiles are artificial in two ways. Firstly, the galaxy
input parameters span a wider range in parameter space than real galaxies. Secondly,
the simulated tiles are significantly more crowded than the actual data itself,
about a factor of 7--8 overdense in galaxies with \sersic\ index $n>2.5$ compared to
a typical \ACS\ image from \gems\ (see Figure \ref{fig_crowding}). They contain many
more LSB galaxies (detected and undetected), adding a complex layer of extra flux to
the background. This makes the simulations more difficult to analyze than real data;
this was intentional since we wanted to push both codes to their limits. In a third
set of simulations we mixed the two types of profile (see \S\ref{sec_results_fake})
to estimate the effects of deblending given a more realistic mix of $n=1$ and $n=4$
galaxies.
To test and compare the two different 2D-fitting routines, the simulated images were
treated as `real' images, i.e. we used exactly the same data pipeline for fitting
that was used for the real \gems\ data analysis. Therefore, all effects which we can
see in the results from simulations should be present in real data as well, although
mixed with many other effects like bulge/disk composite profiles, non-smoothness,
lumpiness and/or spiral features of real galaxies.
\section{Galaxy fitting: Description, Basic Considerations,
Best-fitting Setups} \label{sec_codes}

For the \gems\ analysis, we have used two widely-employed parametric fitting codes
for quantitatively describing galaxy structure and morphology: \galfit\ and
\gimtwod. In this section, we describe both codes and the procedures used to
parametrically fit both the real \gems\ data and the simulations described in the
previous section. The basic considerations for code setup and application to real
data, and the tests which we have performed on simulated data, are useful in general
to other workers in galaxy image fitting. These basic considerations for setup and
application of these (and most) fitting codes are (1) sky estimation, (2) initial
parameter guesses, (3) postage stamp construction, and (4) deblending and/or masking
of neighboring sources. We describe in detail the setups and various tests we
carried out in order to optimize these setups in \S\ref{sec_galfit_general} for
\galfit\ and \S\ref{sec_gimtwod_general} for \gimtwod.

The initial conditions and setups for both \galfit\ and \gimtwod\ are determined
using \sex\ output images and catalogues. We use \sex\ \citep[version 2.2.2
,][]{Bertin} for image parsing and catalog creation. \sex\ detects, deblends,
measures and classifies objects, giving estimates of magnitude, size, $b/a$,
position angle and a star-galaxy classification. In \gems, we found that no single
\sex\ setup satisfactorily detected and deblended both bright, well-resolved
galaxies and faint galaxies near the detection limit. Accordingly, our best setup is
to run \sex\ twice: once to detect the bright objects without splitting them up
(what we call the `cold' version) and once to detect the faint objects (`hot'
version). The two versions are then combined to give one single catalog containing
all objects. The procedure is described in more detail in \citet{Rix} and
\citet{Caldwell}. We do not use the \sex\ output catalogs directly for science;
instead, these values are used as initial estimates for galaxy fitting codes and
their setup. In the following sections we will describe which parameters are taken
as starting guesses and how these values are used for the two parametric galaxy
fitting codes used in this work: \galfit\ and \gimtwod.

\subsection{GALFIT}\label{sec_galfit_general}
\galfit\ is a 2D galaxy fitting software package written by \citet{Peng}. We used
\galfit\ Version 2.0.3b from Feb. 2, 2005 for this analysis. \galfit\ was designed
to extract structural components from galaxy images. Compared to other fitting
techniques it has two main advantages. It uses a Levenberg-Marquardt
downhill-gradient \citep{Press} method to derive the best fit and therefore is
relatively fast, being able to fit roughly 3000 galaxies per day on a dual 2.4 GHz
LINUX processor (when running 4 threads simultaneously to efficiently use all CPU
time). Furthermore, due to its speed and design, it is able to fit an image
containing an arbitrary number of galaxies simultaneously, making it possible to fit
neighboring objects. The main disadvantage of \galfit, in theory, is that it is
possible that it converges on fit solutions that represent a local minimum instead
of giving the global minimum. Our experience with \galfit\ is that in single
component, but multi-object, fits this happens relatively rarely, if at all, both
through the simulations (\S\ref{sec_results_sims}) and through comparison of fitting
results for real galaxies from \galfit\ and \gimtwod\ (\S\ref{sec_results_deep}).

During the fitting process, the model is convolved with a specified PSF to model the
image seeing and then compared to the input image. It is possible to fit the
background sky level during the fitting process, although in this paper we use this
capability for testing purposes only (see \S\ref{sec_galfit_sky}).

In the following section, we will explain the basic setup procedure of \galfit\ in
detail; e.g., cutting postage stamps, estimating the sky background, deciding on how
galaxies should be deblended, and setting up the initial parameters for \galfit. We
developed automated routines for this purpose, and we describe their most important
features in this section. As sky background is of critical importance, we discuss
this issue in some detail in \S\ref{sec_galfit_sky}.

\subsubsection{\galfit\ setup and \galapagos}\label{sec_galfit_setup}
\galfit\ is designed to fit one galaxy of interest at a time. Therefore, we created
an individual postage stamp for each galaxy of interest . These postage stamps were
created, and initial \galfit\ parameter files produced, by an IDL program,
\galapagos\ (\textit{\underline{G}}alaxy \textit{\underline{A}}nalysis over
\textit{\underline{L}}arge \textit{\underline{A}}reas:
\textit{\underline{P}}arameter \textit{\underline{A}}ssessment by
{\sc\textit{\underline{G}}alfit}ting \textit{\underline{O}}bjects from
{\sc\textit{\underline{S}}Extractor}, for further details about \galapagos\ and
details of the procedure see Barden et al., in prep.). For every object in the \sex\
catalog \galapagos\ did the following.
\begin{enumerate}
\item First, \galapagos\ determined the size of the required postage
stamp for each object. This was done using different object sizes and
angles given by \sex:
\begin{equation}
Xsize = 2.5*a*kron*(|\sin(\theta)|+(1-ellip)*|\cos(\theta)|)
\end{equation}
\begin{equation}
Ysize = 2.5*a*kron*(|\cos(\theta)|+(1-ellip)*|\sin(\theta)|)
\end{equation}
where $a$ is the \sex\ output parameter \verb+A_IMAGE+, \verb+kron+ is
\verb+KRON_RADIUS+, $\theta$ is \verb+THETA_IMAGE+ and \verb+ellip+ is
\verb+ELLIPTICITY+. Extensive testing showed that this algorithm for producing
postage stamps was a good compromise between the conflicting needs of having enough
sky pixels present in the postage stamp to give a robust fit of the object, while
keeping the postage stamps small enough to be fit in reasonable amounts of CPU time.

\item In the next step, \galapagos\ decided from this postage stamp and
the aperture map, which secondary objects had to be deblended and fitted
simultaneously and which objects were simply masked out during the fitting process.
For this it created a second map where \sex\ aperture ellipses were increased in
linear size by a factor of 1.5 (a factor of 2.25 larger area). Every object whose
ellipse overlapped with the ellipse of the primary object was fitted simultaneously
using a single \sersic\ profile; every other object with pixels in the postage stamp
was masked out during the fit, using this expanded ellipses as the mask\footnote{In
many other fitting routines the \sex\ segmentation map is used for masking; our
masks are considerably more conservative.}. This way time-consuming fits, with 10 or
more objects to be simultaneously fitted, were avoided in most cases. In total for
around 48\%/31\%/46\% of the fits, at least {\it one} secondary object had to be
taken into account (for $n=1$ simulations, $n=4$ simulations and real galaxies
respectively). In the most crowded situations we find that we needed to
simultaneously fit a maximum of 9/7/12 profiles.

\item After this step the sky background was estimated. For this,
\galapagos\ used the aperture map on the whole science frame (and not only the
postage stamp) and estimated the mean value of all pixels that lay within 6
consecutive elliptical annuli, each with a width of 60 pixels (measured along the
semi-major axis; corresponding to 1.8" using the \gems\ data with 0.03"/pixel).
These 6 annuli partially overlap, with a spacing of 30 pixels  between successive
annuli. The annuli were centered on the primary fitting galaxy (pixels belonging to
a secondary object were ignored in this step). The innermost area is masked out
during this process (the factor of 1.5 magnified aperture ellipse enlarged by a
further 30 pixels). These annuli `marched outward' together in radius in steps of 30
pixels until the gradient of the mean values within the last 6 rings (180 pixels)
was larger than -0.05; the change in the sky value, given that the mean \gems\
F850LP sky background is around 18 counts, was then well below 0.3 \% within this
radial range. The sky was then determined as the mean value of the outermost 6
annuli. This made the area where the sky is determined to be an ellipse between 35
and 215 pixels in semi-major axis for the smallest objects (between 15 and 30
$r_{e}$) and an ellipse of width of 180 pixels at around 4-6 $r_{e}$ for the bigger
objects (for details see Barden et al, 2006, in prep.). We call this sky estimate
the `isophotal sky' in what follows, and testing shows that for fitting with
\galfit\ the `isophotal sky' provides significantly better fitting results than
using sky values from, e.g., \sex\ (see \S\ref{sec_galfit_sky}).

\item In the same step, by dividing the elliptical individual annuli into
8 octants, \galapagos\ was able to detect sky gradients within an annulus as a
function of position angle. Such cases were relatively rare, and were due to nearby
bright objects that did not reach into the postage stamp themselves (especially
bright high \sersic\ index objects with strong wings). \galapagos\ then identified
these objects in the \sex\ catalog automatically and these objects were fitted
simultaneously to eliminate this sky gradient (\galfit\ can fit profiles that are
centered outside of the postage stamp). In the very rare cases that an
identification was not possible although a strong gradient was present (i.e. the
object lay outside of the original \gems\ tile), we fit an artificial object
centered outside the postage stamp in the correct direction to achieve the same
result. In total, 15.2\% of the fits in the simulated disk sample needed an
additional identified profile centered outside of the postage stamp, 1.5\% needed an
artificial, not identified profile (4.3\%, 0.6\% for simulated spheroidal galaxies).
For real galaxies only 3\% of the fits needed an identified object, 0.4\% needed an
artificial profile. Recall that the simulated images contained a large number of
galaxies not recovered by \sex; these galaxies contributed to the background sky
only. These galaxies can lead to `sky' gradients found by \galapagos. This effect
should be, and is, more pronounced in the sample where fewer galaxies are recovered.
\begin{deluxetable*}{ll}
\tablewidth{0pt} \tablenum{3} \tabletypesize{\scriptsize}
\tablecolumns{2} \tablecaption{Starting guesses for \galfit\ when using
\galapagos} \tablehead{\colhead{Parameter} & \colhead{Starting guess from
\sex}} \startdata
mag      & \verb+MAG_BEST+\\
$r_{e}$  & $ 0.162 \cdot \verb+FLUX_RADIUS+^{1.87}$ \\
$b/a$       & 1 - \verb+ELLIPTICITY+ \\
PA       & \verb+THETA_IMAGE+ \\
$n$        & 1.5 \\
x,y  & the postage stamp is centered on the primary object\\
 & positions of secondary objects can be derived from \sex\\
\enddata \label{tab_start_guess_galfit}
\end{deluxetable*}
\item The last step for setting up \galfit\ was the determination of the
starting guesses for the different fitting parameters from \sex\ and writing them to
a \galfit\ start file automatically (see Table \ref{tab_start_guess_galfit}). We
decided to fit single \sersic\ profiles to all galaxies (with a starting value of
1.5 for the \sersic\ index). Starting magnitudes were given by \sex\
\verb+MAG_BEST+, sizes were given using \verb+FLUX_RADIUS+ (we used the formula
$r_{e}= 0.162 \cdot R_{flux}^{1.87}$, where $R_{flux}$ is \verb+FLUX_RADIUS+. This
formula was determined empirically using simulations). The axis ratio $b/a$ was
derived by taking the \sex\ \verb+ELLIPTICITY+, the position angle by
\verb+THETA_IMAGE+. Furthermore the position of the objects within its postage stamp
was required, which was directly given by the cutting process of the postage stamps
(the object is centered within its postage stamp, see step 1). The parameter
diskiness/boxiness in \galfit\, was {\it fixed} to 0 (no boxiness/diskiness) for all
our fits. Furthermore, as described above, the estimated sky value from step 3 {\it
was held fixed during the fit}. Each object that had to be deblended during the
fitting process was included (from step 2) with its appropriate starting values; all
other objects were masked out (using a mask image with the by a factor of 1.5
enlarged \sex\ apertures that tells \galfit\ which pixels it should use and which
pixels it should ignore during the fit). Finally, the \gems\ PSF (see Jahnke et al.,
in prep.) was provided to \galfit.
\end{enumerate}

We adopted a set of fitting constraints for \galfit\ which prevented the code from
exploring unphysical (and time-consuming) areas of parameter space. We used
$0.2<n<8$, $0.3<r_{e}<500$ [pixels of 0.03$"$ in size] and fixed the fit magnitude
to be within 5 mag of the \sex\ \verb+MAG_BEST+.

We used exactly the same constraints for real galaxies, also using the same setup
procedure. Whenever we state that \galfit\ fitted `successfully', we mean that
\galfit\ returned a result (it did not crash during the fit) {\it and} the fit did
not run into any of the constraints given above.

\subsubsection{\galfit\ sky test}\label{sec_galfit_sky}
The estimate of the sky background is of critical importance in determining
parametric or non-parametric descriptions of galaxy surface brightness profiles
\citep[e.g.,][]{deJong1996}. While in principle it is possible to fit the sky level
as an extra parameter, such a procedure requires that the surface brightness profile
being used is an {\it accurate} description of the real galaxy light profiles. An
alternative is to estimate the sky level as carefully as possible prior to the fit
and hold it fixed.
\begin{figure*}[htb]
\begin{center}
\includegraphics[width=8cm,angle=0]{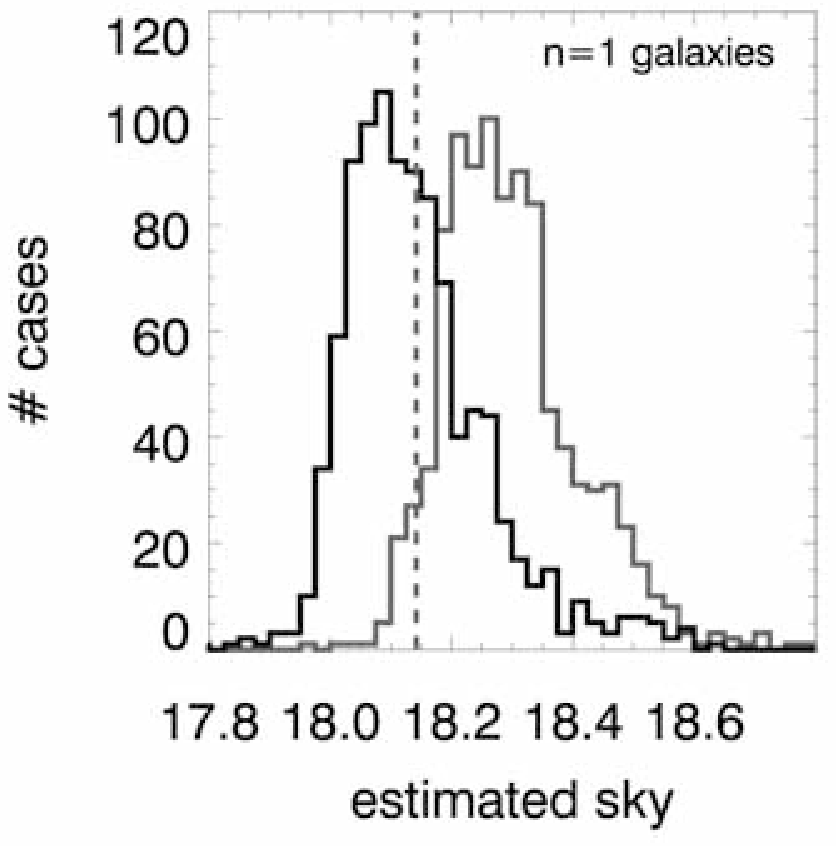}
\includegraphics[width=8cm,angle=0]{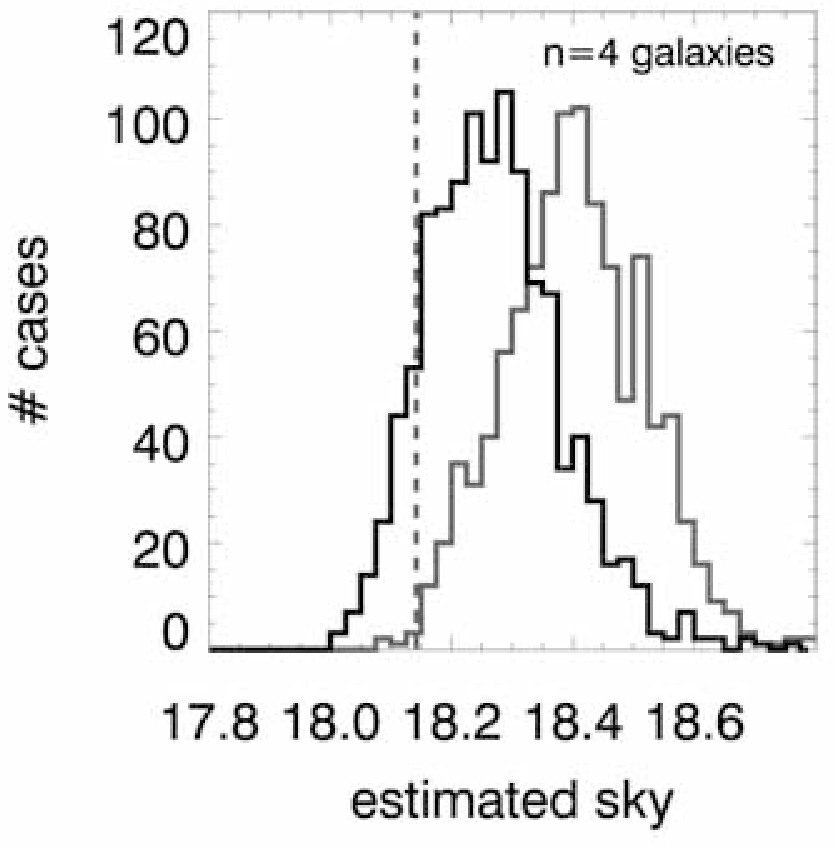}
\end{center}
\caption{This plot shows the recovered sky values for both used estimators, \sex\
(grey) and isophotal (determined within \galapagos, black) for each of the simulated
samples. The vertical dashed line indicates the true value. One can clearly see that
both methods tended to overestimate the sky value for the spheroid sample, mainly
due to the large wings of galaxies in this sample contributing to the sky level.
\label{fig_sky_values}}
\end{figure*}
In this section we quantify the effect of different assumptions/estimates of the sky
level for \galfit\ (the results of the equivalent test for \gimtwod\ are shown in
\S\ref{sec_gimtwod_sky}). We test three setups: {\it i)} the isophotal sky, {\it
ii)} the sky value determined by \sex, and {\it iii)} allowing sky to be a free
parameter, to be estimated by \galfit.

In Figure \ref{fig_sky_values} one can see the difference between the sky values
derived by the two sky estimation methods, \galapagos\ and \sex, for the two
different samples of simulated $n$=1 and $n$=4 galaxies, respectively. Because the
simulations were added to a sky frame composed of empty patches of real sky, the
{\it true} sky values were known to be $18.14 \pm 0.03$, indicated by the vertical
dashed line in both plots. \sex\ recovers a mean value of 18.29 ($\sigma$=0.10) for
disk galaxies and 18.40 ($\sigma$=0.11) for spheroidal galaxies. The isophotal
estimator in \galapagos\ gives a mean value of 18.13 ($\sigma$=0.10) for disk
galaxies and 18.26 ($\sigma$=0.11) for spheroidal galaxies. Although all
distributions have around the same width, one can see that both methods recover the
sky better for the low \sersic-index sample. Furthermore, in both samples, the
isophotal estimator gives back rather more accurate sky values.

That \sex\ recovers a sky value that is slightly too high has been noted before --
e.g., by the \goods\ team\footnote{see
$http://www.stsci.edu/science/goods/catalogs/r1.0z\_readme/$, chapter 5.1 Local sky
background} and was the reason why we decided to write our own isophotal sky
estimator.

That the sky is easier to estimate for the $n=1$ simulations than for the $n=4$
simulations can be partly explained by our simulation of a number of large, low
surface brightness galaxies which escape detection by \sex\ and which inflate the
sky surface brightness. Since galaxies with high \sersic\ index $n$ have more
extended wings the effect of contamination in the outskirts is larger for $n=4$
simulations than for $n=1$. There is a further effect for $n=4$ galaxies: since the
sky estimates provided by \sex\ and \galapagos\ only probe out to $<6 r_e$ for
brighter galaxies, there is a residual contribution to the sky from the galaxy
itself which becomes more serious as $n$ increases.
\begin{figure*}[htb]
\begin{center}
\includegraphics[width=12cm,angle=0]{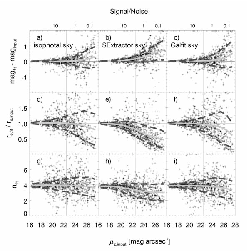}
\end{center}
\caption{This plot shows the fitting results of the spheroid galaxy sample when
fitted with \galfit\ using three different sky estimates (see
\S\ref{sec_galfit_sky}): isophotal sky from \galapagos\ (left), \sex\ sky (middle)
and the sky returned by \galfit\ when allowed to fit it as a free parameter. The
X-axis shows the simulated mean surface brightness within $r_{e}$ defined by $\mu =
mag + 2.5\cdot \log(2\cdot b/a \cdot \pi\cdot r_{e}^{2})$ , where mag is the
magnitude, $b/a$ the axis ratio and $r_{e}$ the half-light radius of the object in
arcsec. The thin vertical grey line in the plot indicates the brightness of the sky
background. The upper X-axis shows the mean signal-to-noise ratio per pixel within
$r_{e}$ calculated by: $S/N = \langle\varrho\rangle \cdot [\langle\varrho\rangle +
\langle\varrho_{sky}\rangle + \sigma_{sky}]^{-1/2}$ , where $\langle\varrho\rangle$
is the average counts in a galaxy pixels within $r_{e}$ (basically $\mu$),
$\langle\varrho_{sky}\rangle$ is the background flux within a pixel and
$\sigma_{sky}$ is the uncertainty of the background sky estimation. Although this
number is only a rough approximation, it gives a feeling about the mean $S/N$ of the
galaxies. The Y-axis shows magnitude difference (fitted - simulated), size ratio
(fitted/simulated) and \sersic\ index fitting results. Perfect parameter recovery is
indicated as the horizontal thin dark-grey line. The thick light-grey line and the
thick dashed dark-grey line indicate the mean value and $1\sigma$ values for
different surface brightness bins. The small crosses show the galaxies that where
fitted `successfully', meaning that the fit returned a result and that it did not
run into fitting constraints. As one can clearly see getting a good estimation of
the sky level is important. Both using the isophotal sky estimation (leftmost
column) and using the sky level as a free parameter during the fit (rightmost
column) return more reliably results than using \sex\ sky estimations.
\label{fig_sky_comp_bulge}}
\end{figure*}
\begin{deluxetable*}{lcccccccccccc}
\tablewidth{0pt} \tablenum{4} \tabletypesize{\scriptsize} \tablecolumns{11}
\tablecaption{\gimtwod: Fitting of $n=4$ Simulations using \galfit: Bright subsample
with $\mu_{\rm in}<22.5$ and ${\rm mag}_{\rm in}<22.5$} \tablehead{\colhead{Sky
used} & \multicolumn{2}{c}{\sersic\ $n$} & \multicolumn{2}{c}{$r_{50}$ ratio} &
\multicolumn{2}{c}{$\Delta{\rm mag}$} & \multicolumn{2}{c}{$e$ ratio} &
\multicolumn{2}{c}{$\Delta{\rm PA}$} & \colhead{Quality}\\
\cline{2-11} \\
\colhead{} & \colhead{mean} & \colhead{$\sigma$} & \colhead{mean} &
\colhead{$\sigma$} & \colhead{mean} & \colhead{$\sigma$} & \colhead{mean} &
\colhead{$\sigma$} & \colhead{mean} & \colhead{$\sigma$} & \colhead{}} \startdata
isophotal sky  &  3.99 &  0.27 &  1.00 &  0.05 & 0.00 &  0.03 &  0.99 &  0.04 & 0.17 & 1.54 & 0.06 \\
\sex\ sky  &  3.79 &  0.29 &  0.96 & 0.07 & 0.03 & 0.04 & 1.00 & 0.04 & 0.17 & 1.56 & 4.66 \\
\galfit\ sky  & 3.94 & 0.24 & 0.99 & 0.06 & 0.01 & 0.03 & 1.00 & 0.04 & 0.16 & 1.55 & 0.38 \\
\enddata
\tablecomments{This Table summarizes the results from using different sky estimators
with \galfit\ for bright galaxies; see Table \ref{tab_galfit_sky_faint} for results
for faint galaxies.
\\
The columns give deviations (resistant mean values clipped at $3 \sigma$) from the
simulated value and scatter for the 5 key fitting parameters. The $\sigma$ values
given are values computed iteratively for all galaxies within 3 $\sigma$.\\
The last column gives the fit quality. This number is defined as:
\begin{equation}
Quality = 1000*[(\Delta n/4-1)^2 + (\Delta re-1)^{2} + (\Delta mag)^2 + (\Delta
(b/a)-1)^2 + (\Delta PA/180)^2)]
\end{equation}
where $\Delta$ values are given as the mean values in the table. This quantity is a
fairly intuitive combination of the different fit parameters, indicating in broad
terms which setups perform well (low values) and which perform poorly (high values).
One can see that indeed using the isophotal sky as given by \galapagos\ and using
the sky level as a free parameter during the fit return much more reliable results
than the \sex\ sky already for these bright galaxies. Using the isophotal sky seems
to be the ideal setup. \label{tab_galfit_sky_bright}}
\end{deluxetable*}
\begin{deluxetable*}{lcccccccccccc}
\tablewidth{0pt} \tablenum{5} \tabletypesize{\scriptsize} \tablecolumns{12}
\tablecaption{\gimtwod: Fitting of $n=4$ Simulations using \galfit: Faint subsample
with $23.5<\mu_{\rm in}<26.0$} \tablehead{\colhead{Sky used} &
\multicolumn{2}{c}{\sersic\ $n$} & \multicolumn{2}{c}{$r_{50}$ ratio} &
\multicolumn{2}{c}{$\Delta{\rm mag}$} & \multicolumn{2}{c}{$e$ ratio} &
\multicolumn{2}{c}{$\Delta{\rm PA}$} & \colhead{Quality}\\
\cline{2-11} \\
\colhead{} & \colhead{mean} & \colhead{$\sigma$} & \colhead{mean} &
\colhead{$\sigma$} & \colhead{mean} & \colhead{$\sigma$} & \colhead{mean} &
\colhead{$\sigma$} & \colhead{mean} & \colhead{$\sigma$} & \colhead{}} \startdata
isophotal sky  & 3.95 & 1.13 & 0.97 & 0.44 & 0.02 & 0.42 & 1.04 & 0.18 & 0.20 & 6.31 & 2.92\\
\sex\ sky  & 3.05 & 0.98 & 0.61 & 0.27 & 0.41 & 0.39 & 1.05 & 0.18 & -0.06 & 6.38 & 375.33 \\
\galfit\ sky  & 3.78 & 1.16 & 0.94 & 0.47 & 0.09 & 0.44 & 1.04 & 0.19 & 0.07 & 6.12 & 16.93\\
\enddata
\tablecomments{Same as Table \ref{tab_galfit_sky_bright}, but for faint galaxies. As
one can clearly see from this table and as was expected, fitting faint galaxies is
more difficult. Using the isophotal sky during the fit returns the best fitting
results, slightly better than the internal estimation in \galfit. Using the \sex\
sky returns significantly worse results. \label{tab_galfit_sky_faint}}
\end{deluxetable*}

We compare the fitting results with the three different sky setups in Figure
\ref{fig_sky_comp_bulge}. We only show results for the sample of simulated $n=4$
galaxies; the results for the $n=1$ galaxies were qualitatively similar but the
systematic effects are much weaker, showing very little difference between the three
different sky setups. The Y-axis shows the deviation of the three key parameters
magnitude, $r_e$ and $n$ from their true values, and the X-axis shows the simulated
mean surface brightness $\mu_{input}$ of the galaxies within an ellipse with
semi-major axis $r_{e}$ and the axis ratio $b/a$:
\begin{equation}
\mu = mag + 2.5\cdot \log(2\cdot b/a \cdot \pi\cdot r_{e}^{2})
\end{equation}
where mag is the magnitude, $b/a$ the axis ratio and $r_{e}$ the half-light radius
of the object in arcsec. The factor of two accounts for the fact that only half the
light is within the half-light radius. The top axis shows the mean $S/N$ per pixel
corresponding to that average surface brightness $\mu$, given by
\begin{equation}
S/N = \langle\varrho\rangle \cdot [\langle\varrho\rangle +
\langle\varrho_{sky}\rangle + \sigma_{sky}]^{-1/2}
\end{equation}
where $\langle\varrho\rangle$ is the average countrate [in $e^{-}$] for galaxy
pixels within $r_{e}$, $\langle\varrho_{sky}\rangle$ is the background flux [in
$e^{-}$] within a pixel, and $\sigma_{sky}$ is the uncertainty of the background sky
estimate, obtained from the empty sky image.

At faint surface brightness levels, one can see that magnitudes are typically
overestimated (i.e., are recovered too faint), sizes are systematically
underestimated, and \sersic\ indices are typically underestimated (Figure
\ref{fig_sky_comp_bulge}). The effects are subtle and affect only galaxies much
fainter than the sky surface brightness for the isophotal sky and the sky fit by
\galfit. These effects set in at much higher surface brightness (approximately 2
\magarc\ above the sky level) for \sex-derived sky values.

Following these test results (Tables \ref{tab_galfit_sky_bright} and
\ref{tab_galfit_sky_faint}), and the general concern that galaxies may deviate from
parametric descriptions of their light profiles in their outer parts, we choose to
use the isophotal sky estimate for \galfit\ analysis for this paper (and those used
in other papers, e.g., \citealp{Barden2005}). Should one not have access to accurate
sky values from \galapagos\ or a similar routine, the test results show also that
allowing \galfit\ to estimate the sky levels for single \sersic\ profiles is an
acceptable alternative, provided that the surface brightness profiles of the
galaxies of interest are well-approximated by a \sersic\ profile over a wide range
of radii.

\subsection{GIM2D} \label{sec_gimtwod_general}
\gimtwod\ (\textit{\underline{G}}alaxy \textit{\underline{Im}}age
\textit{\underline{2D}}) was written by Luc Simard \citep{Simard98, Simard02} as an
IRAF package for the quantitative morphological analysis of galaxies. We use version
3.1 for the analysis in this paper. For a single \sersic\ fit we work in
7-dimensions with the bulge fraction parameter set to $B/T=1$; thus, we find the
best-fit model described by $f_{\rm tot},r_{\rm e},e,\phi_B,dx,dy$, and $n$. During
the fit, the images are deconvolved with a given PSF. \gimtwod\ uses the Metropolis
algorithm to find a $\chi^2$ minimum, which makes it less prone to settle on local
minima. On the other hand, this algorithm is time consuming. Accordingly, to process
large datasets, \gimtwod\ ought to be run on many machines in parallel.

\subsubsection{GIM2D setup}\label{sec_gimtwod_setup}
As with \galfit, \gimtwod\ requires certain generic considerations for galaxy
profile fitting: (1) postage stamp construction, (2) nearby companion masking, (3)
background sky estimation, and (4) initial parameter guesses. We did not use
\galapagos\ to set up \gimtwod's galaxy fit for two reasons: \gimtwod\ is embedded
into IRAF whereas \galapagos\ requires IDL; and the simultaneous fitting of galaxies
is not supported in \gimtwod, whereas much of \galapagos's algorithm is devoted to
making decisions about which galaxies are to be simultaneously fit. Therefore,
\gimtwod\ is set up by using a different procedure, which we describe in this
section.

Starting with the combined hot/cold \sex\ output catalogues, a square postage stamp
was cut from the large image, centered on each galaxy with size given by $4a_{\rm
iso}\times 4a_{\rm iso}$, where $a_{\rm iso}$ is the major axis diameter of the
\sex\ isophotal area in pixels (the minimum postage stamp size we allowed was
$101\times101$ pixels). \gimtwod\ masks out nearby objects using \sex\ segmentation
maps: discussion of the consequences of this procedure is presented later in this
section, and in \S\ref{sec_results_sims_bulge}. For sky estimation and defining the
best part of the fitting parameter space to explore, \gimtwod\ has several important
setup parameters that allow the user to modify its behavior. In this section, we
describe some of the most important ones -- parameters that we find to critically
affect the performance of the code.

The parameter \verb+`dobkg'+ specifies whether \gimtwod\ determines the background
itself (\verb+`dobkg'+$=$yes) or fixes the sky to a user-defined value
(\verb+`dobkg'+$=$no). With \verb+`dobkg'+$=$yes, \gimtwod\ calculates the
background prior to galaxy fitting directly from the postage stamp images of each
source using only non-object (sky) pixels as specified by the \sex\ segmentation
map. As such, this method is closely dependent upon extracting a large enough image
to get a reliable sky measurement. Once determined, the sky value is held fixed
during the fitting. If \verb+`dobkg'+$=$no, \gimtwod\ assumes that the postage
stamps have background equal to zero; therefore, the user may use an external method
to estimate the sky and subtract this from the input images. \gimtwod\ does offer an
option to fit the background offset (parameter $db$) as a free parameter during
fitting, but this is not recommended when working with real galaxies with
non-idealized profiles. We test the effect of different sky estimates in detail in
\S\ref{sec_gimtwod_sky}.

\gimtwod, like \galfit, has constraints which can be applied to limit the regions of
parameter space searched for solutions. \gimtwod\ starts with a user-specified
parameter space, given by the initial value and minimum/maximum hard limits for each
parameter to be fit. \gimtwod\ has an option to automatically narrow the focus of
the input parameter space by setting \verb+`initparam'+$=$yes. With this setup
option \gimtwod\ uses FOCAS-like image moments based on information extracted from
the \sex-created segmentation map to estimate the hard limits for the model
parameter space.

Under all setups \gimtwod\ starts in the Initial Condition Finder (ICF) mode, which
explores the user-specified parameter space coarsely to find the best initial model
guess. In practice, the ICF creates $N_{\rm ICF}$ models throughout the allowed
parameter space, selects the best one, and then reduces the search volume by a
factor equal to $N_{\rm ICF}$. The final result from the ICF is used as the starting
point by the Metropolis algorithm. The \gimtwod\ website gives a default value of
$N_{\rm ICF}=100$.

To find the {\it best-fitting} setup we rigorously tested a large number of
different setups of \gimtwod. We do not discuss all of the different setups here;
the most important ones are shown in Tables \ref{tab_gimtwod_bright} (for bright
galaxies) and \ref{tab_gimtwod_faint} (for faint galaxies) and will be discussed in
detail in the following sections starting with the recommended \gimtwod\ setup
(\S\ref{sec_gimtwod_recom}), sky tests (\S\ref{sec_gimtwod_sky}), other tests
(\S\ref{sec_gimtwod_tests}), concluding with the final adopted best-fitting setup
(\S\ref{sec_gimtwod_best}).

\subsubsection{\gimtwod\ recommended setup} \label{sec_gimtwod_recom}
\begin{figure*}[htb]
\begin{center}
\includegraphics[width=12cm,angle=0]{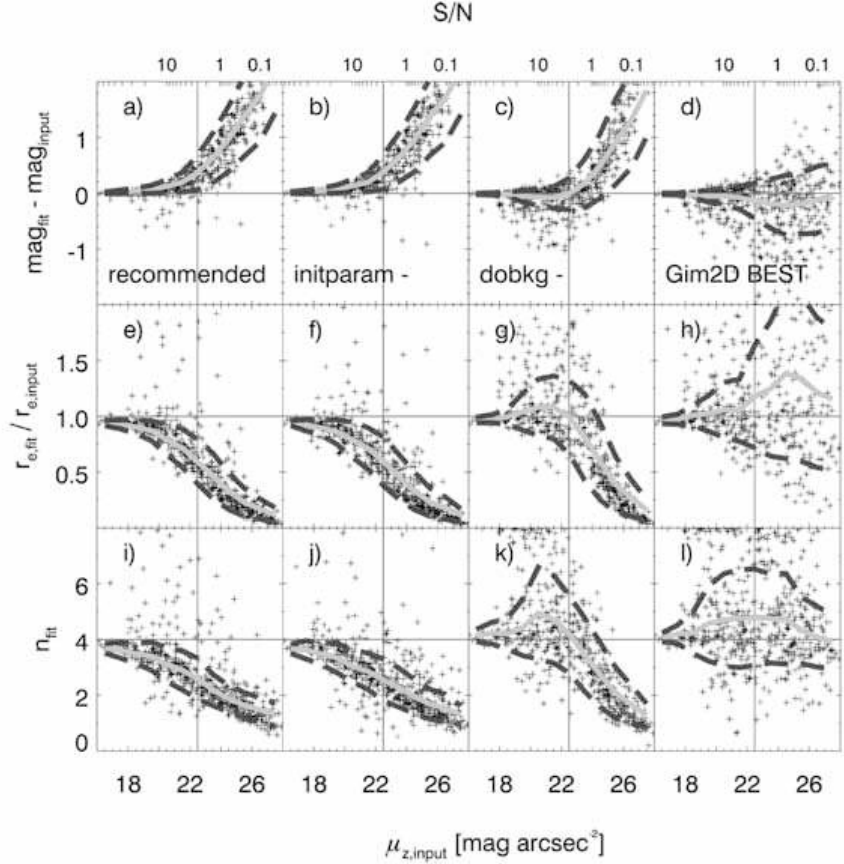}
\end{center}
\caption{This plot compares the recommended \gimtwod\ setup to setups where we used
different settings of the \gimtwod\ parameters `initparam' and `dobkg' [from left to
right: setup K (initparams=+, dobkg=+, recommended),J (-, +),C (+, -)and A (-, -;
best), see Table \ref{tab_gimtwod_bright}]. The X-axis shows the simulated surface
brightness of the galaxies. The Y-axes again show magnitude difference, size ratio
and \sersic\ index fitting results. The thick light-grey line and the thick dashed
dark-grey line indicate the mean value and the $1\sigma$ line for different surface
brightness bins. One can see easily that our {\it best-fitting} \gimtwod\ setup (see
\S\ref{sec_gimtwod_best}) fits galaxies with much less systematic bias than the
initial setup recommended on the \gimtwod\ webpage. This is especially true for
galaxies fainter than the surface brightness of the sky. \label{fig_gim2d_comp}}
\end{figure*}

In Figure \ref{fig_gim2d_comp}, in the leftmost panels, we show fitting results for
the setup that is recommended on the \gimtwod\ webpage\footnote{The GIMFIT2D
description is http://www.hia-iha.nrc-cnrc.gc.ca/STAFF/lsd/gim2d/, and specifies the
last program update of March 19, 2001} (setup K in Table \ref{tab_gimtwod_bright}).
This recommended setup, in particular, has \verb+`dobkg+'$=$yes and
\verb+`initparams'+$=$yes; i.e., \gimtwod\ determines the sky level and fitting
constraints from \sex\ output. As is clear from this plot, this setup produces
unsatisfactory results even for fairly high surface brightness galaxies and where
\gems\ survey completeness is still quite high. The systematic errors are already
$\sim 50$\% in $r_e$ near the sky level. Fitting results are strongly systematically
biased towards fainter magnitudes, smaller sizes and lower concentrations.

As most galaxy surveys aim to push their analysis down to faint levels, the ideal
performance of any fitting code is to provide parameter estimates that are free of
significant systematic trends. Therefore, we deem the recommended setup to not be
suitable for the \gems\ survey. In an attempt to improve the \gimtwod\ performance,
we tried a number of different strategies, among them different settings of
\verb+`initparams'+  and \verb+`dobkg'+.

Through extensive testing, we find that the best results are obtained when {\it
both} \verb+`initparams'+$=$no and \verb+`dobkg'+$=$no (see rightmost panels in
Figure \ref{fig_gim2d_comp}), and when the \sex\ local background is used (see
\S\ref{sec_gimtwod_sky}). Setting \verb+`initparams'+$=$no and dobkg$=$yes produces
very modest improvement. Setting \verb+`dobkg'+$=$no and initparam$=$yes helps
considerably, giving satisfactory results for galaxies with surface brightness
higher than the sky surface brightness\footnote{This is the setup that was used by
\citet{McIntosh2005} for their study of the evolution of the early-type $n>2.5$
galaxy luminosity--size and stellar mass--size relations. Their sample of $n>2.5$
galaxies all had F850LP surface brightness brighter than 22.5 mag\,arcsec$^{-2}$,
and inspection of the third row of panels in Fig.\ \ref{fig_gim2d_comp} and setup J
in Table \ref{tab_gimtwod_bright} shows that at these limits the \gimtwod\ fitting
results suffer from $\la 10$\% biases.}. As can be seen in Fig.
\ref{fig_gim2d_comp}, using a fixed background (\verb+dobkg'+$=$no) and setting
\verb+initparams'+$=$no removes the strong systematic trend towards poorer fits for
low surface brightness galaxies, albeit with large scatter at the faint end. As
explained in \S \ref{sec_gimtwod_setup}, the \verb+initparams'+$=$no option allows
\gimtwod\ to explore the full range of parameter space when determining the best-fit
solution. In contrast, setting this option to 'yes' narrowly constrains the
magnitude and $r_{e}$ for objects where the \sex\ segmentation map severely misses
the total extent of galaxies below the sky brightness. It is worth noting that
compared to these two parameter choices, other effects such as the precise fixed sky
value used (see \S\ref{sec_galfit_sky}) and the image size
(\S\ref{sec_gimtwod_tests}) appear to produce only minor improvements.
\begin{deluxetable}{lll}
\tablewidth{0pt} \tablenum{6} \tabletypesize{\scriptsize} \tablecolumns{3}
\tablecaption{Parameter limits used for \gimtwod\ when using initparam=no}
\tablehead{\colhead{Parameter} & \colhead{min} & \colhead{max}} \startdata
mag         & 20 & 27\\
$r_{e}$     & 0.3 & 500 \\
ellipticity & 0.0 & 1.0\\
PA          & all & all \\
$n$         & 0.2 & 8.0 \\
centering x & 0 & 3.0 \\
centering y & 0 & 3.0 \\
\enddata \label{tab_start_guess_gim2d}
\end{deluxetable}
The minimum and maximum limits of the parameter space that we allowed in our best
setup are manually set to span more than the entire range of the simulations in
terms of size, luminosity and \sersic\ index, more than the physically useful
parameter range of real galaxies, so that the solutions are not `pinned up' against
the boundary values artificially imposed on them (the actual values are given in
Table \ref{tab_start_guess_gim2d}). Fits that ran into any of the given fitting
constraints were removed from the sample for the analysis in this paper.

From our findings, we {\it strongly recommend} that \gimtwod\ users {\it avoid} the
dobkg=yes option and {\it be cautious} of the surface brightness effects that arise
when using initparam=yes.

\subsubsection{\gimtwod\ sky test} \label{sec_gimtwod_sky}
\begin{figure*}[htb]
\begin{center}
\includegraphics[width=16cm,angle=0]{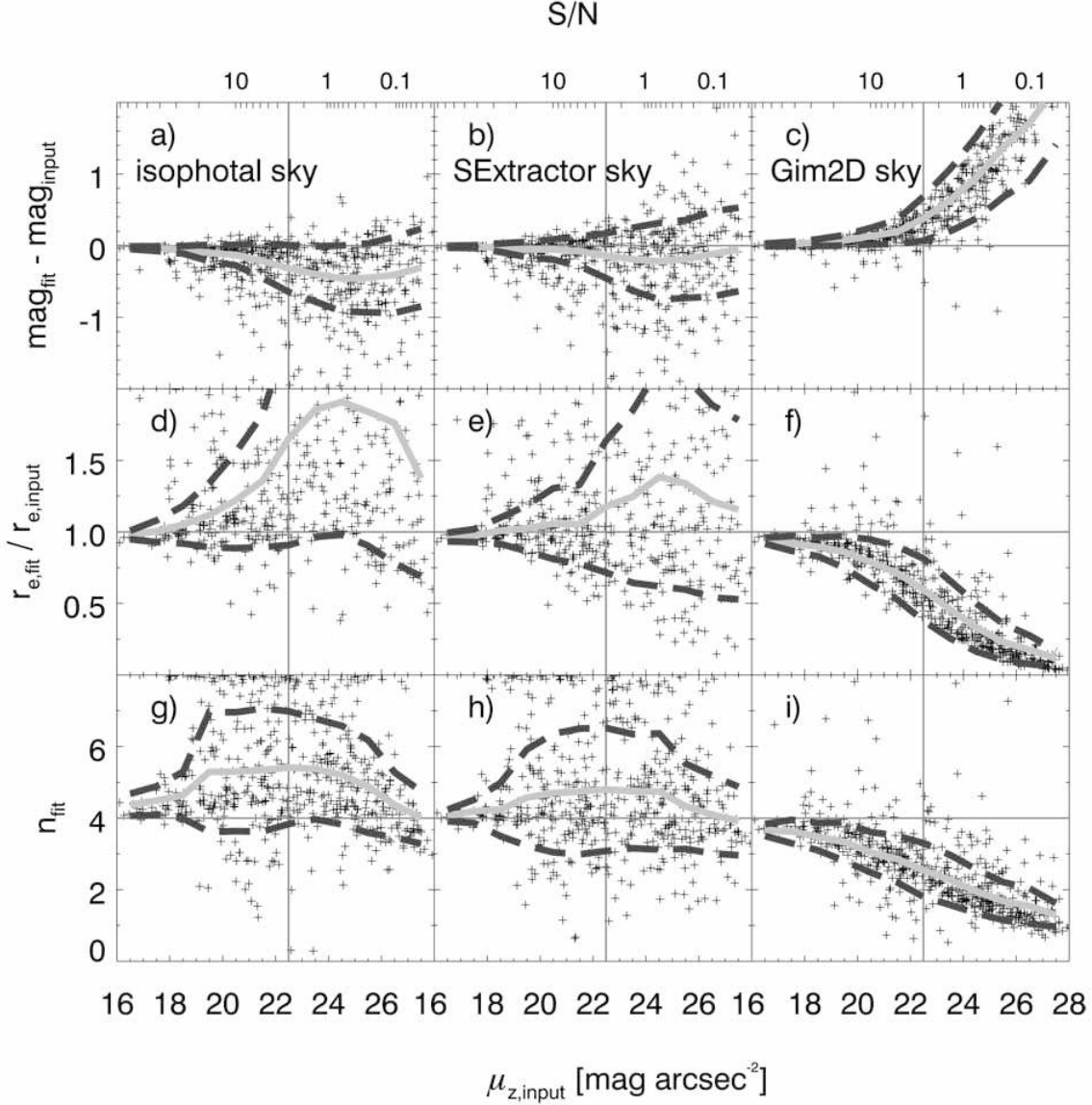}
\end{center}
\caption{This plot shows the same as Figure \ref{fig_sky_comp_bulge} but for
\gimtwod\ results (setups F, A and J in Table \ref{tab_gimtwod_bright}). As is
immediately obvious from these plots, \gimtwod\ performs best when using the \sex\
background held fixed during the fit. \label{fig_sky_comp_gimtwod_bulge}}
\end{figure*}

In the above, we showed that sky value estimation can dramatically affect \gimtwod\
fits using the default (recommended) setting. In this section, we repeat the sky
analysis for \gimtwod\ as carried out in \S\ref{sec_galfit_sky} for \galfit. We
tested \gimtwod\ using \sex\ local sky (setups A, B and C in Table
\ref{tab_gimtwod_bright}) and the isophotal sky that we used for \galfit\ (setups D,
E, F, G and H). We carried out one test -- setup I -- where we used the `real'
background of 18.14 counts, determined on the `sky' image used in the simulation
process. Such a setup is of academic interest only, as for real galaxies it is
impossible to measure such a sky value. Nonetheless, this test gives insight into
the performance of \gimtwod\ when the actual, known, sky value is used as an input
for galaxy fitting. For setup J and K we tried fixing the sky background to the
value determined directly by \gimtwod\ (\verb+`dobkg'+=yes). We show results from
these different tests in Figure \ref{fig_sky_comp_gimtwod_bulge}.
It is interesting that \gimtwod\ performs somewhat better using \sex\ local sky,
whereas we showed in \S\ref{sec_galfit_sky} that \galfit\ performs somewhat better
using the isophotal sky from \galapagos. It is likely that the cause of this
behavior is related to how \galfit\ and \gimtwod\ deal with nearby neighbors. Since
\galfit\ simultaneously fits neighboring galaxies which overlap with the galaxy of
interest, the isophotal sky estimate better represents the background pedestal that
is common to the neighboring sources. \gimtwod, on the other hand, is unable to
simultaneously fit neighbors, and relies on masking neighbors using the \sex\
segmentation map. Thus the `effective sky' for \gimtwod\ includes flux from the
outer parts of the galaxy itself and neighboring sources; as the \sex\ sky is
derived from the same `sky' area used for fitting, it is a more appropriate value.

\subsubsection{\gimtwod\ other tests} \label{sec_gimtwod_tests}
\begin{deluxetable*}{lcccccccccccc}
\tablewidth{0pt} \tablenum{7} \tabletypesize{\scriptsize} \tablecolumns{12}
\tablecaption{\gimtwod: Fitting of $n=4$ Simulations using \gimtwod: Bright
subsample with $\mu_{\rm in}<22.5$ and ${\rm mag}_{\rm in}<22.5$}
\tablehead{\colhead{Setup} & \colhead{$N/N_{\rm tot}$} & \multicolumn{2}{c}{\sersic\
$n$} & \multicolumn{2}{c}{$r_{50}$ ratio} & \multicolumn{2}{c}{$\Delta{\rm mag}$} &
\multicolumn{2}{c}{$e$ ratio} &
\multicolumn{2}{c}{$\Delta{\rm PA}$} & \colhead{Quality}\\
\cline{3-12} \\
\colhead{} & \colhead{} & \colhead{mean} & \colhead{$\sigma$} &
\colhead{mean} & \colhead{$\sigma$} & \colhead{mean} & \colhead{$\sigma$}
& \colhead{mean} & \colhead{$\sigma$} & \colhead{mean} &
\colhead{$\sigma$} & \colhead{}} \startdata
 A  & 164/540 &  4.34 &  0.61 &  1.02 &  0.13 & -0.04 &  0.07 &  1.01 &  0.05 &  0.00 &  2.2 &    7.02 \\
 B  & 168/533 &  4.37 &  0.43 &  1.02 &  0.09 & -0.03 &  0.05 &  1.01 &  0.06 & -0.16 &  1.5 &    8.38 \\
 C  & 165/549 &  4.32 &  0.56 &  1.01 &  0.10 & -0.03 &  0.07 &  1.02 &  0.06 & -0.10 &  1.8 &    6.31 \\
 D  & 161/533 &  5.57 &  1.49 &  1.14 &  0.22 & -0.14 &  0.13 &  1.02 &  0.07 & -0.20 &  2.0 &  154.57 \\
 E  & 164/531 &  4.84 &  0.76 &  1.11 &  0.17 & -0.09 &  0.08 &  1.01 &  0.06 & -0.17 &  2.0 &   44.42 \\
 F  & 165/539 &  4.75 &  0.67 &  1.10 &  0.17 & -0.08 &  0.08 &  1.01 &  0.06 &  0.05 &  2.0 &   35.63 \\
 G  & 162/539 &  4.82 &  0.75 &  1.10 &  0.17 & -0.08 &  0.07 &  1.01 &  0.06 &  0.34 &  2.2 &   42.02 \\
 H  & 161/545 &  4.81 &  0.76 &  1.12 &  0.16 & -0.09 &  0.09 &  1.01 &  0.06 & -0.03 &  1.6 &   40.63 \\
 I  & 163/533 &  5.68 &  1.40 &  1.27 &  0.30 & -0.16 &  0.14 &  1.01 &  0.06 & -0.03 &  1.7 &  177.28 \\
 J  & 167/551 &  3.30 &  0.49 &  0.82 &  0.14 &  0.09 &  0.09 &  1.00 &  0.05 & -0.18 &  1.9 &   30.40 \\
 K  & 168/546 &  3.33 &  0.51 &  0.83 &  0.13 &  0.09 &  0.09 &  1.01 &  0.05 & -0.01 &  1.6 &   28.32 \\\enddata
\tablecomments{This Table summarizes the results from all \gimtwod\
testing for bright galaxies; see Table \ref{tab_gimtwod_faint} for
results for faint galaxies.
\\
$N/N_{tot}$ gives the numbers of galaxies $N$ selected from the total sample of
$N_{tot}$ that \gimtwod\ returns a result for each setup. The following columns give
deviations (resistant mean values clipped at $3 \sigma$) from the simulated value
and scatter for the 5 key fitting parameters. The $\sigma$ values given are values
computed iteratively for
all galaxies within 3 $\sigma$.\\
The last column gives the fit quality as defined in Table
\ref{tab_galfit_sky_bright}.\\
Explanation of the setups: \\
(A) SExtr. local bkg, {\it initparam=no}, $N_{\rm ICF}=100$,  $4a_{\rm iso}$ image sizes, {\it best} setup\\
(B) SExtr. local bkg, {\it initparam=no}, $N_{\rm ICF}=100$,  $2a_{\rm iso}$ image sizes\\
(C) SExtr. local bkg, {\it initparam=yes}, $N_{\rm ICF}=100$,  $4a_{\rm iso}$ image sizes\\
(D) isoph. bkg, {\it initparam=no}, $N_{\rm ICF}=100$,  $6a_{\rm iso}$ image sizes\\
(E) isoph. bkg, {\it initparam=no}, $N_{\rm ICF}=25$, and $4a_{\rm iso}$ image sizes\\
(F) isoph. bkg, {\it initparam=no}, $N_{\rm ICF}=100$,  $4a_{\rm iso}$ image sizes\\
(G) isoph. bkg, {\it initparam=no},$N_{\rm ICF}=400$, and $4a_{\rm iso}$ image sizes\\
(H) isoph. bkg, {\it initparam=yes}, $N_{\rm ICF}=100$, $4a_{\rm iso}$ image sizes\\
(I) ${\rm bkg}=18.14$, {\it initparam=no}, $N_{\rm ICF}=100$,  $4a_{\rm iso}$ image sizes\\
(J) {\it dobkg=yes}, {\it initparam=no}, $N_{\rm ICF}=100$,  $4a_{\rm iso}$ image sizes\\
(K)  {\it dobkg=yes}, {\it initparam=yes}, $N_{\rm ICF}=100$,  $4a_{\rm iso}$ image sizes, {\it recommended} setup\\
\label{tab_gimtwod_bright}}
\end{deluxetable*}
\begin{deluxetable*}{lcccccccccccc}
\tablewidth{0pt} \tablenum{8} \tabletypesize{\scriptsize} \tablecolumns{12}
\tablecaption{\gimtwod: Fitting of $n=4$ Simulations using \gimtwod: Faint subsample
with $23.5<\mu_{\rm in}<26.0$} \tablehead{\colhead{Setup} & \colhead{$N/N_{\rm
tot}$} & \multicolumn{2}{c}{\sersic\ $n$} & \multicolumn{2}{c}{$r_{50}$ ratio} &
\multicolumn{2}{c}{$\Delta{\rm mag}$} & \multicolumn{2}{c}{$e$ ratio} &
\multicolumn{2}{c}{$\Delta{\rm PA}$} & \colhead{Quality}\\
\cline{3-12} \\
\colhead{} & \colhead{} & \colhead{mean} & \colhead{$\sigma$} &
\colhead{mean} & \colhead{$\sigma$} & \colhead{mean} & \colhead{$\sigma$}
& \colhead{mean} & \colhead{$\sigma$} & \colhead{mean} &
\colhead{$\sigma$} & \colhead{}} \startdata
  A & 151/540 &  4.63 &  1.45 &  1.38 &  0.78 & -0.20 &  0.55 &  1.05 &  0.23 & -1.10 &  7.4 &   24.83 \\
  B & 149/533 &  4.52 &  0.92 &  1.44 &  0.67 & -0.23 &  0.43 &  1.05 &  0.21 & -0.13 &  7.0 &   17.35 \\
  C & 154/549 &  2.46 &  0.86 &  0.50 &  0.23 &  0.57 &  0.48 &  1.01 &  0.19 & -0.64 &  5.5 &  149.08 \\
  D & 147/533 &  5.52 &  1.36 &  3.22 &  1.93 & -0.76 &  0.57 &  1.00 &  0.20 & -1.22 & 10.6 &  148.96 \\
  E & 147/531 &  5.05 &  1.27 &  1.89 &  0.95 & -0.47 &  0.49 &  1.02 &  0.16 & -1.17 &  7.6 &   69.38 \\
  F & 149/539 &  5.13 &  1.26 &  1.94 &  0.93 & -0.47 &  0.46 &  1.03 &  0.20 & -1.00 &  6.3 &   80.32 \\
  G & 148/539 &  5.24 &  1.30 &  2.36 &  1.45 & -0.52 &  0.51 &  1.02 &  0.16 & -1.43 &  6.5 &   97.78 \\
  H & 153/545 &  2.45 &  0.92 &  0.53 &  0.27 &  0.53 &  0.50 &  0.96 &  0.16 & -0.88 &  8.1 &  150.72 \\
  I & 146/533 &  5.38 &  1.19 &  2.84 &  1.51 & -0.73 &  0.53 &  0.99 &  0.25 & -0.88 &  7.0 &  122.88 \\
  J & 155/551 &  1.79 &  0.44 &  0.28 &  0.11 &  1.08 &  0.49 &  1.02 &  0.19 & -0.57 &  9.2 &  306.48 \\
  K & 154/546 &  1.78 &  0.40 &  0.28 &  0.12 &  1.10 &  0.47 &  1.04 &  0.15 & -1.86 &  7.5 &  310.23 \\
\enddata
\label{tab_gimtwod_faint}
\end{deluxetable*}

To determine the best-fitting setup to use with \gimtwod, we performed 11 different
tests (rows A-K) as shown in Tables \ref{tab_gimtwod_bright} (bright galaxies) and
\ref{tab_gimtwod_faint} (faint galaxies). For the `bright' galaxies, we selected all
$N$ galaxies with $\mu_{sim}<22.5$ \magarc\ and $mag_{sim}<22.5$ (representing the
sample of early-type galaxies from \citet{McIntosh2005}, i.e., those important for
surveys of early-type galaxy evolution) from the set of $N_{\rm tot}$ galaxies in
the sample for which \gimtwod\ returned a result. The `faint' sample included
galaxies with $23.5 <\mu_{sim}<26$ \magarc. We then calculated the mean of the
recovered value or ratios of the different fit parameters and the 68\% confidence
interval.

In our visual examination of the properties of the outliers in these distributions,
we found that most of the non-Poisson scatter is caused by contamination of the
outer isophotes of the object of interest by nearby neighbors. While this issue is
discussed in more detail later in \S \ref{sec_results_deblending}, we illustrate
this behavior by running \gimtwod\ on 3 different postage stamp widths of $2a_{\rm
iso}$ (setup B), $4a_{\rm iso}$ (setup A), and $6a_{\rm iso}$ (setup D). These tests
find that there is an increase in scatter for larger image size, consistent with the
expectation of contamination. Using $2a_{iso}$ reduces the extreme outlier fraction
somewhat from $4a_{iso}$. Yet, since such outliers are a small fraction of the
objects, this change in postage stamp size had relatively little impact on the RMS
scatter (see tables \ref{tab_gimtwod_bright} and \ref{tab_gimtwod_faint}). The
best-fit stamp cutouts we adopt here have sides equal to $4a_{\rm iso}$. This seems
to be the best compromise between a postage stamp large enough so that \gimtwod\
includes enough of the important outskirts of the galaxies for fitting, but small
enough that neighboring galaxies are reasonably rare and CPU requirements are
reasonable. For comparison, the stamp sizes used in \galfit\ fits are nearly always
larger in area than $4a_{\rm iso}\times 4a_{\rm iso}$, due to the requirement of
simultaneously fitting neighboring galaxies. We will further quantify the effect of
neighboring galaxies in \S\ref{sec_results_sims_bulge} using the set of simulated
spheroids examined there.

We also tested whether the initial number of ICF models affected our fitting
results. Holding all other setup choices constant we compared the results from fits
to the $n=4$ simulations with the default value of $N_{\rm ICF}=100$ (setup F in
Table \ref{tab_gimtwod_bright}), to results for $N_{\rm ICF}=25$ (setup E) and 400
(setup G). We found that the results are independent of the number of ICF models.

\subsubsection{\gimtwod\ best-fitting setup} \label{sec_gimtwod_best}
As is apparent especially from Table \ref{tab_gimtwod_faint} and Figure
\ref{fig_gim2d_comp}, the best combination of parameters for our simulations was
given by setup A, using \sex\ background ({\it dobkg=no}), {\it initparam=no} and
$4a_{\rm iso}$ as image sizes. We choose this setup to be our {\it best} and use it
throughout this paper to compare \gimtwod\ results with \galfit\ results.

\section{GALFIT/GIM2D Comparison Using Optimized Setups}\label{sec_results}
In this section, we discuss the results of testing our best setups of \gimtwod\ and
\galfit. Section \ref{sec_results_sims} describes the results obtained using the
simulated images with artificial distributions of galaxy parameters as explained in
\S\ref{sec_sims}. Section \ref{sec_results_fake} describes a very similar test using
simulated galaxies having more realistic parameter distributions, as derived from
real galaxies recovered from individual \gems\ survey fields. Section
\ref{sec_results_deep} sums up the results of tests where real images of different
depths were fit and the results intercompared.

\subsection{Results of Fitting Simulated Galaxy Images}\label{sec_results_sims}
\subsubsection{Results of pure disk simulations} \label{sec_results_sims_disk}
\begin{figure*}[htb]
\begin{center}
\includegraphics[width=12cm,angle=0]{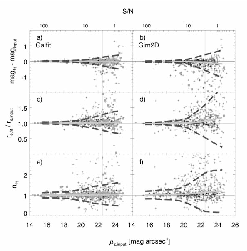}
\end{center}
\caption{Fitting results for \galfit\ (left) and \gimtwod\ (right) for the set of
simulated $n=1$ galaxies. The X-axis again shows the input surface brightness, the
thin vertical grey line indicates the brightness of the sky background. The Y-axes
are the same as in Figure \ref{fig_sky_comp_bulge}. The input value is indicated as
the horizontal thin dark-grey line, in this case representing a \sersic\ index of 1
for the sample of disk galaxies. The mean value of the deviations and a $3\sigma$
line are indicated for different surface brightness bins. The small crosses show the
galaxies that where fitted `successfully' by both codes. Grey squares indicate
galaxies that where fitted by one code only and the other code did not return a
meaningful result. The quality parameters as defined in Table
\ref{tab_galfit_sky_bright} for this set of galaxies is 0.05/5.27 for \galfit\
(bright/faint) and 1.01/251.67 for \gimtwod\ showing that \galfit\ returns more
reliable results for simulated disk galaxies in its optimal setup than \gimtwod.
\label{fig_disk_sims}}
\end{figure*}
Figure \ref{fig_disk_sims} shows both \galfit\ and \gimtwod\ results for the set of
simulated disk galaxies with an exponential $n=1$ light profile. Of the 1600
galaxies simulated in this sample, 997 (62\%) were recovered by \sex. Of these, 979
(98\%) were successfully fitted by \galfit, 12 (1.2\%) ran into constraints, 6
(0.6\%) fits crashed. \gimtwod\ fitted 870 (87\%) successfully, 46 (5\%) ran into
fitting constraints, 81 (8\%) of the fits crashed. There are 4 (0.4\%) galaxies for
which both codes failed.

Crosses in Figure \ref{fig_disk_sims} represent galaxies that were fitted by both
codes. Grey squares show galaxies that were fitted by that code only; the other code
failed to return a useful result either through running into one of the fitting
constraints, or the fit crashed. The thick light-grey line and the thick dashed
dark-grey line indicate the mean value and the $3\sigma$ lines for different surface
brightness bins of all galaxies that were fitted using that code. The left column
shows fitting results using \galfit, the right column shows the results of the same
set of simulations using \gimtwod. The X-axis, showing the simulated surface
brightness of the galaxies, is the same for all 6 plots. The 3 rows show the results
for magnitude (plots a and b), size (c and d) and \sersic\ index (e and f),
respectively. The thin vertical line indicates the brightness of the sky background
within the \gems\ survey. This is also roughly the limit up to which real galaxies
are used for science within \gems. The Y-axes show deviations of the fitting values
to the true parameter values, the horizontal thin line indicating the ideal value,
which, in case of this galaxy sample, is simply the simulated value.

We will discuss here and in all other sections the behavior of the codes in 3
different surface brightness bins: firstly, the galaxies of highest surface
brightness which one clearly would want to be fitted well with any code; secondly,
galaxies within a surface brightness bin of 1 magnitude around the surface
brightness of the sky; and thirdly, the faintest galaxies, much fainter than the sky
surface brightness. The third are the galaxies that are obviously hardest to fit.
Here the results from the two codes differ the most from each other.

To summarize our general findings, for $n$=1 galaxies brighter than the sky's
surface brightness, there is no significant mean offset between the input and
recovered values in Figure \ref{fig_disk_sims}; however, the scatter in the
\gimtwod\ results is somewhat larger. For this sample of galaxies, this would mean
that the final results would be statistically unaffected by one's choice of fitting
code, but for individual objects the reliability of the \galfit\ results is slightly
higher.

For galaxies around the sky surface brightness, there are small systematic trends
and increased scatter for our setup of \gimtwod: a size ratio of 1.06
($r_{fitted}/r_{simulated}$, $\sigma\approx0.18$) for \gimtwod\, and a ratio of 1.02
for \galfit\ ($\sigma\approx0.08$). This trend continues towards fainter surface
brightness, although at no point does the systematic size offset exceed 20\%.

From Figure \ref{fig_disk_sims} (grey squares show objects fitted only by the
respective code), one can easily see that \galfit\ returns a result more often than
\gimtwod, although the fraction of galaxies with failed fits is small in both cases.
It is interesting to note that the properties of galaxies with failed fits is
somewhat different between the two codes: galaxies not fit by \galfit\ (those fit
only by \gimtwod) are fainter than average, the parameters are discrepant even using
\gimtwod, whereas galaxies not fit by \gimtwod\ (those fit only by \galfit) are fit
almost as well by \galfit\ as other galaxies with the same surface brightness.

\subsubsection{Results of pure spheroid simulations} \label{sec_results_sims_bulge}
\begin{figure*}[htb]
\begin{center}
\includegraphics[width=12cm,angle=0]{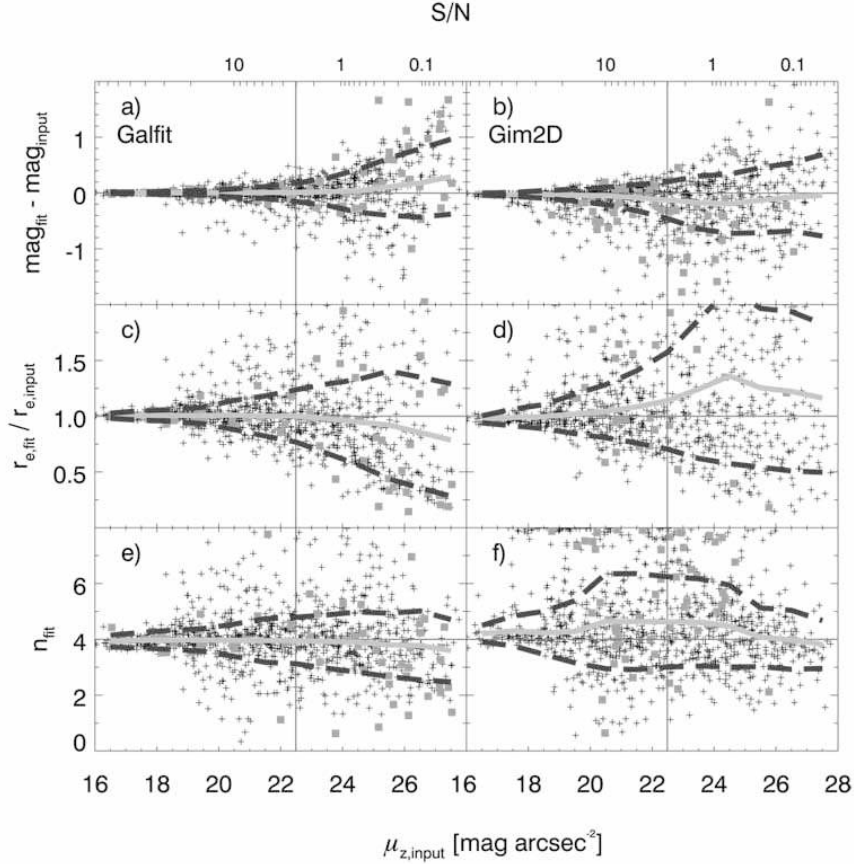}
\end{center}
\caption{Code comparison for $n=4$ galaxies. This figure is formatted in a similar
way to Figure \ref{fig_disk_sims}, but for the sample of simulated $n=4$ galaxies.
The X-axis is shifted by 2 \magarc\ compared to Fig \ref{fig_disk_sims};
furthermore, in this plot the grey dashed line represent the $1\sigma$ limits. The
value for $\sigma$ in this sample is around 3 times as large as was the case for
disk galaxies. The true value of 4 for the \sersic\ index is again indicated as the
horizontal thin line. The quality parameter for bright/faint galaxies is 0.06/2.92
for \galfit\ and 8.09/122.00 for \gimtwod\ using this sample of galaxies. Numbers
given here are different from the numbers given in Tables \ref{tab_gimtwod_bright}
and \ref{tab_gimtwod_faint} as a different, larger sample of galaxies was used for
this analysis. From fewer galaxies in Table \ref{tab_gimtwod_faint}, the best setup
A there returns a slightly higher mean \sersic\ index which translates into a higher
quality parameter. The numbers show clearly that \galfit\ returns better results
than \gimtwod. Interestingly, the quality parameter for \gimtwod\ is smaller for
faint spheroids than for faint disks, suggesting that these galaxies are fit more
reliably. This might be due to the way the quality parameter was calculated in
detail (using resistant mean offset values cut at 3 $\sigma$ which removes large
scatter, indeed the systematic offset is smaller in case of spheroidal galaxies, the
scatter does not go into the quality number) but does not reflect the plots, the
disk plots look better due to smaller scatter. \label{fig_bulge_sims}}
\end{figure*}

Figure \ref{fig_bulge_sims} shows the same plots as Figure \ref{fig_disk_sims} but
for the simulated set of $n=4$ profiles representing the light profile of a typical
early-type galaxy. The total number of galaxies recovered in this sample out of 1600
simulated objects was 1091 (68\%). Of these, only 2 (0.2\%) crashed in \galfit, 56
(5.1\%) ran into constraints; \gimtwod\ crashed on 31 galaxies (2.8\%), 36 (3.3\%)
additional fits ran into fitting constraints. 54 (5.0\%) galaxies were fitted by
\gimtwod\ that were not fitted by \galfit, 63 (5.8\%) galaxies were fitted by
\galfit\ and not fitted by \gimtwod, both codes crashed on 4 (0.4\%) galaxies in
common.
When comparing Figure \ref{fig_bulge_sims} to Figure \ref{fig_disk_sims}, one should
be aware that the $X$-axis is shifted by 2 \magarc\ towards fainter surface
brightness, and that the dashed lines for clarity indicate $1\sigma$ instead of
$3\sigma$ as in Figure \ref{fig_disk_sims}. It is clear that both codes recover the
parameter values for $n=4$ galaxies significantly less accurately than was the case
for the $n=1$ disks, resulting in a substantially larger scatter. This is due to two
different effects. Firstly, spheroidal profiles are in principle harder to fit due
to the importance of the outskirts of the light profile -- this makes using an
appropriate sky estimate much more important for a successful fit. Secondly, due to
the large amount of light in the faint wings of the galaxies, neighboring objects
have a much bigger influence on the fit of the galaxy of interest than was the case
for the exponential light profiles. This effect is particularly important for this
simulated galaxy sample, because it was designed to have an unrealistically high
number of large $n=4$ galaxies.

As was the case for disk galaxies, both codes are basically indistinguishable for
high surface brightness galaxies in a statistical sense. For galaxies with surface
brightness close to that of the sky, our implementation of \galfit\ recovers
slightly better parameter values than \gimtwod\ (size ratio of 1.00,
$\sigma\approx0.23$ and a somewhat asymmetric error distribution for \galfit; size
ratio of 1.14, $\sigma\approx0.44$ and more asymmetric errors for \gimtwod). The
trend continues towards lower surface brightness, with the \gimtwod\ showing
increasingly important systematic offsets and a substantially increased scatter. The
directionality and asymmetry of the scatter in all plots (\gimtwod\ and \galfit) are
caused by neighboring contamination that is not fully removed, keeping in mind that
32\% of the simulated galaxies escape detection by \sex\ because of their low
surface brightness.

\subsubsection{Deblending effects} \label{sec_results_deblending}
Given the significant differences in philosophy when it comes to the deblending
techniques between \galfit\ (multiobject fitting \& masking) and \gimtwod\ (masking
only), we explore the recovery of input parameters as a function of the immediate
environment of a galaxy for both codes. We analyze the subset of 390 (out of a total
of 1033) $n=4$ simulated galaxies where \galapagos\ decided that \galfit\ needed to
simultaneously fit two or more profiles. This has the advantage that only
significant neighbors are included in this analysis and should be sufficient to
demonstrate the magnitude of the influence of deblending on the quality of galaxy
fitting with \galfit\ and \gimtwod.

\begin{figure*}[htb]
\begin{center}
\includegraphics[width=12cm,angle=0]{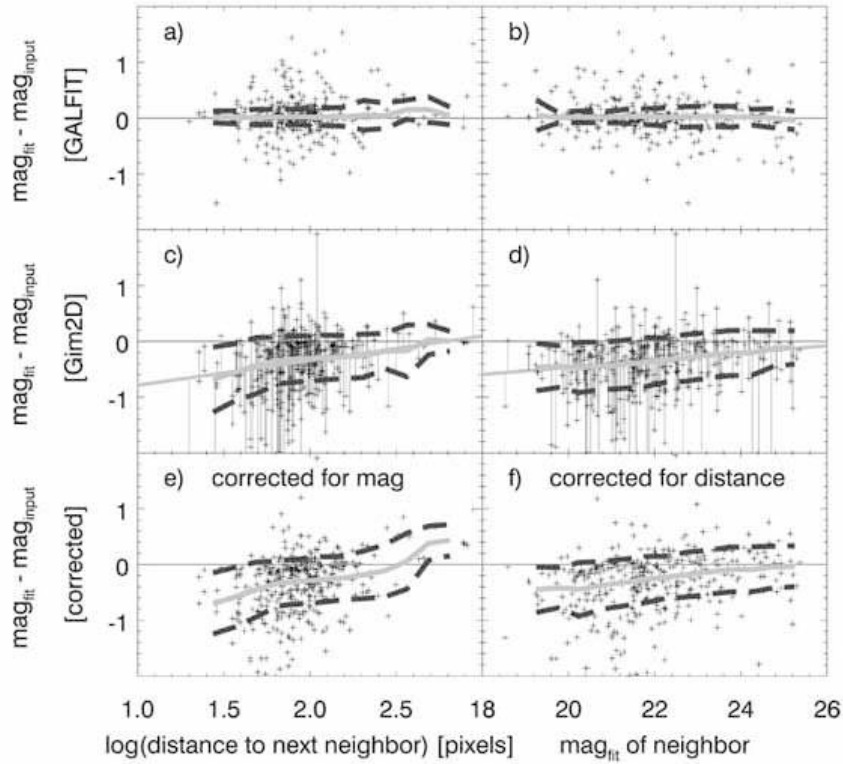}
\end{center}
\caption{The impact of neighboring galaxies on fit results with \galfit\ (upper row)
and \gimtwod\ (middle and lower row). The left column shows magnitude deviations
from the simulated values as a function of the distance to the next neighbor; the
right column shows the difference between the recovered and simulated values as a
function of the brightness of the nearest neighbor. \gimtwod\ shows strong
systematic offsets as a function of both distance to the nearest neighbor and its
brightness. In the lower two panels, we try to correct for the systematics observed
in panels c and d by showing the distance dependence of the offset-magnitude
relation residuals (panel e), and the magnitude dependence of the offset-distance
relation residuals (panel f). \label{fig_neighbor}}
\end{figure*}

The results are summarized in Figure \ref{fig_neighbor}, showing the difference
between recovered and input magnitude as a function of the distance to the next
neighbor (left) and as a function of the brightness of this neighbor (right).
Fitting neighboring objects simultaneously, \galfit\ (panels a and b) is able to
deblend these galaxies reliably, and the deviations of the fitting magnitudes is
independent of both distance and brightness of the closest neighbor. For \gimtwod\
(panels c and d), it is clear that fitting residual is a strong function of both
distance and brightness of the nearest neighbor. The closer and brighter a
neighboring object is, the larger is the magnitude deviation. In an attempt to
disentangle the influence of distance and brightness, we try to correct for the
systematics observed in panels c and d by removing the offsets and the slope,
showing the results in panels e and f. It is clear that distance and brightness
effects of nearest neighbor cannot be easily corrected, thus can significantly
impact the performance of \gimtwod\ in recovering the true parameters for simulated
$n=4$ galaxies. For isolated galaxies, \gimtwod\ does an excellent job of recovering
the properties of $n=4$ galaxies.

\subsection{Results of Simulations representing simulated GEMS tiles} \label{sec_results_fake}
Bearing in mind the importance of neighboring galaxies in determining the quality of
fit, we repeated the above analysis using a sample of galaxies where $n=1$ and $n=4$
galaxies were intermixed with realistic clustering, sizes and magnitudes. Towards
this goal, simulations were produced from the \galfit\ results of two real \gems\
tiles using recovered values of magnitude, position and size. The only parameter
that was changed was the \sersic\ index. Every galaxy with a real \sersic\ index of
2.5 or smaller was simulated with a \sersic\ index of 1; all others with a \sersic\
index of 4. These simulations have the advantage that they are better able to
estimate the uncertainties of galaxy fits with \gems\ data.

\begin{figure*}[htb]
\begin{center}
\includegraphics[width=12cm,angle=0]{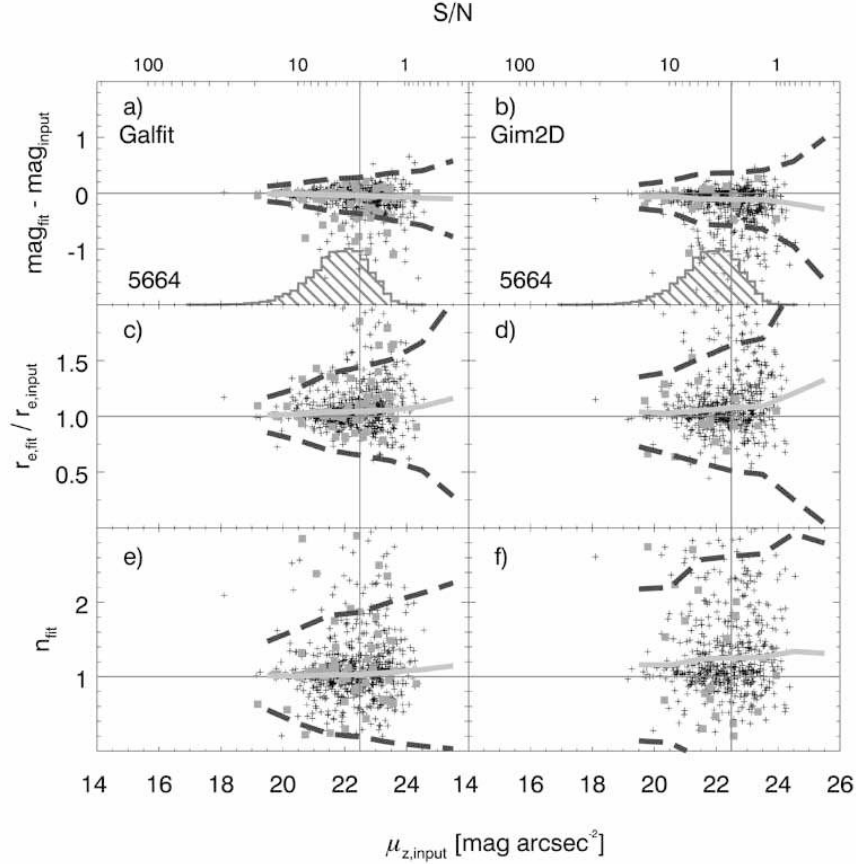}
\end{center}
\caption{Fit results from re-simulated \gems\ tiles (see \S\ref{sec_results_fake}),
disk galaxies. Same as Fig.\ \ref{fig_disk_sims}, but for a sample of 2 \gems\ tiles
that were re-simulated in order to create a more realistic distribution of galaxy
parameters and object density (the results for the $n=4$ galaxies in this sample are
shown in Figure \ref{fig_gems_fake_bulge}). In the upper panel we overplot the
surface brightness histogram of the 5664 disk galaxies that where selected for
analysis by \citet{Barden2005} showing where fitting accuracy is especially
important. \label{fig_gems_fake_disk}}
\end{figure*}

\begin{figure*}[htb]
\begin{center}
\includegraphics[width=12cm,angle=0]{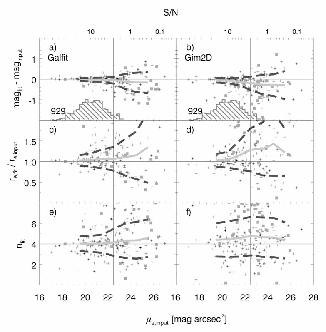}
\end{center}
\caption{Results from re-simulated \gems\ tiles, $n=4$ galaxies. Same as Fig.\
\ref{fig_bulge_sims}, but for the $n=4$ galaxies in the re-simulated \gems\ tiles.
The histogram in the upper panel shows the surface brightness distribution of the
929 red-sequence galaxies that were selected for analysis in \citet{McIntosh2005}.
\label{fig_gems_fake_bulge}}
\end{figure*}

The results are shown in Figures \ref{fig_gems_fake_disk} and
\ref{fig_gems_fake_bulge}. It is worth noting that the range in galaxy surface
brightness is much smaller in these simulations, although we have left the $X$-Axis
the same as in the previous plots to facilitate comparison with these. We also show
the surface brightness histograms of galaxies used in Barden et al. (2005; disk
galaxies) and McIntosh et al. (2005; spheroidal galaxies), to show which areas of
parameter space are especially important for scientific analysis of data.
Inspecting Figs.\ \ref{fig_gems_fake_disk} and \ref{fig_gems_fake_bulge}, it becomes
clear that \gimtwod\ and \galfit\ perform more similarly for galaxy populations with
clustering and properties typical of medium-depth cosmological \HST\ surveys than
for purely simulated data. \galfit\ shows increased scatter and mild systematic
offsets compared to the earlier simulations. In the case of the $n=1$ galaxies the
difference in behavior is especially pronounced: it is clear that the presence of
realistically clustered $n=4$ galaxies around $n$=1 galaxies is a larger source of
random error in galaxy fitting for both \galfit\ and \gimtwod\ than in pure $n$=1
simulations. \gimtwod\ shows very similar behavior compared to the earlier
simulations, with still larger scatter and systematic offsets than \galfit.
\subsection{Results of deep-shallow tests using GOODS and GEMS data}
\label{sec_results_deep}
\begin{figure*}[htb]
\begin{center}
\includegraphics[width=12cm,angle=0]{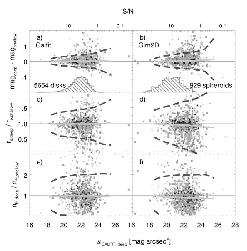}
\end{center}
\caption{Comparison of fits to deep vs. shallow images. The left column shows
results using \galfit\ and the right column shows results for \gimtwod. Both codes
were run on the same sample of real galaxies. Note that in this plot the X-axis
shows the \galfit\ surface brightness derived from the deeper \goods\ data. The
Y-axis shows the deviations of the three key parameters of the galaxies between the
`deep' and the `shallow' fit. In the uppermost plots we again overplotted the
histograms of the disk galaxy sample of \citet[][left histogram]{Barden2005} and the
spheroid-dominated sample of \citet[][right histogram]{McIntosh2005}, right
histogram) to highlight out the area of parameter space where fitting and
independence of the image depth is particularly important. \label{fig_deep_shallow}}
\end{figure*}
Simulations have the disadvantage that the galaxies have unrealistically simple
structure and light-profiles that are known {\it a priori} to be the same as the
profiles used for fitting. Accordingly, in this section we test the performance of
the codes on real galaxies. This goal is not straightforward to achieve, inasmuch as
one does not know what the real parameters of a given galaxy are, or indeed whether
or not real galaxies are well described by the \sersic\ light-profile that was used
during our analysis. Instead, we take an empirical approach and test whether the
fitting results obtained for real galaxies are sensitive to the image depth by
comparing fitting results from the same galaxies in the 1-orbit depth \gems\ survey
and the overlapping 5-orbit depth \goods\ survey. If they were sensitive to the
image depth, it would show that the \sersic\ profile is of limited applicability in
describing the light profile of real galaxies.

Inspection of Figure \ref{fig_deep_shallow} shows clearly that {\it both} codes are
reasonably self-consistent when fitting the same galaxies on images of different
depth, i.e. neither \galfit\ nor \gimtwod\ depends strongly on image depth. While
robustness to image depth does not imply that the fitting results are necessarily
correct, it does give confidence that issues such as low surface brightness disks
missing from shallow \HST\ imaging, departures from \sersic\ profiles at fainter
surface brightness levels, etc., do not appear to seriously compromise the
reliability of fitting parameters in 1-orbit depth \HST/\ACS\ data.

\subsection{Error estimations from GIM2D and GALFIT}\label{sec_error_bars}
\begin{figure*}[htb]
\begin{center}
\includegraphics[width=12cm,angle=0]{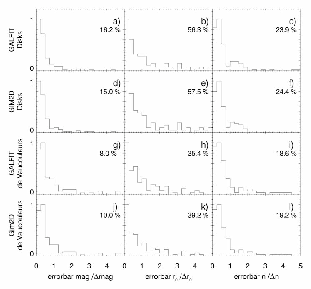}
\end{center}
\caption{A comparison of estimated and real errors for two simulated samples of disk
($n=1$) galaxies and spheroidal ($n=4$) galaxies. Shown is the histogram of the
errorbars $\sigma$ devided by the deviation from the input value $\Delta$. We show
$\sigma$/$\Delta$ instead of the more intuitive quantity $\Delta$/$\sigma$, which
would show a very wide distribution and effects are not as obvious as here.
Calculated as $\sigma$/$\Delta$, in principle for 68\% of all galaxies this value
should be $>$1. The number given in each plot shows the fraction of galaxies for
which this is true. One can easily see that magnitude and \sersic\ index errorbars
are dramatically underestimated by both \galfit\ and \gimtwod; the $r_{e}$
uncertainties are significantly better estimated. \label{fig_errorbars}}
\end{figure*}
It is interesting to consider if the internal error estimates from \galfit\ and
\gimtwod\ are reasonable reflections of the more realistic uncertainties given by
how well the codes recover input parameters for simulated galaxies. In Fig.\
\ref{fig_errorbars}, we address this issue by exploring the distribution of the
error estimate $\sigma$ divided by the deviation of the fit result from the true
value $\Delta$. One can see a strong peak of values with $\sigma/\Delta \ll 1$,
i.e., for these galaxies the deviation $\Delta$ is much larger than the error
estimate $\sigma$\footnote{This behavior was the motivation for plotting
$\sigma$/$\Delta$ instead of the more intuitive quantity $\Delta$/$\sigma$.}. Under
the assumption that the error estimates are correct, $\sigma / \Delta$ should be
$>$1 for 68\% of the galaxies. Fig.\ \ref{fig_errorbars} shows that $\sigma / \Delta
> 1$ for much less than 50\% of the cases; i.e., both \galfit\ and \gimtwod\
substantially underestimate the true fit uncertainties, clearly indicating that the
dominant contribution to fitting uncertainty is not shot and read noise; instead
contamination by neighbors, structure in the sky, correlated pixels, profile
mismatch, etc., dominate the errors. Fig.\ \ref{fig_errorbars} shows no difference
between the histograms of $\sigma / \Delta$ for \galfit\ and \gimtwod; i.e.,
\galfit\ and \gimtwod\ both underestimate the errors to a similar extent.
Accordingly, in this work and all other \gems\ works, we have {\it not} used the
error estimates given by \galfit\ or \gimtwod\ on an object-by-object basis, relying
instead on the mean and width of the parameter distributions from Figs.\
\ref{fig_gems_fake_disk} and \ref{fig_bulge_sims} at the surface brightness of the
galaxy in question.

The uncertainties given in Table \ref{tab_fit_results} are calculated and
interpolated by using the surface brightness $\mu$ and the \sersic\ index $n$ and
the results from the simulated data (see \S\ref{sec_gems_results} for details about
this procedure).

\subsection{Further considerations}\label{sec_consider}
In the course of our preparation of \citet{Barden2005} and \citet{McIntosh2005}, we
found that there were two additional practical considerations that potential users
of \galfit\ and \gimtwod\ may wish to consider.
\begin{itemize}
\item \galfit\ is substantially less CPU intensive than \gimtwod,
reducing the cost and time of fitting large datasets.

\item \gimtwod, at least in our  implementation, failed to return a
fitting result reasonably frequently, requiring manual intervention to restart the
code. When trying to fit large datasets, we found this to be labor-intensive. In
contrast, \galfit\ treated each fit as an individual task and therefore was run from
shell scripts one fit after the other; if \galfit\ does not return a fitting result,
the script automatically starts the next fit, requiring no interaction from the
user.
\end{itemize}

\section{Comparison with Pignatelli et al.\ (2006)} \label{sec_gasphot}
In this paper we present an extensive and thorough test of the two different 2-D
galaxy fitting codes \galfit\ and \gimtwod. In this section, we compare our results
with \citet{Pignatelli}, who compared results from testing these two codes with
their own 1-dimensional profile fitter, \gasphot.

Pignatelli et al. conclude that \gasphot\ performed substantially better for
significantly (realistically) blended objects than either \galfit\ or \gimtwod. In
the course of our testing, we found a number of differences between our analysis and
\citet{Pignatelli}, which we felt may significantly affect their conclusions.
\begin{itemize}
\item For the simulations examined in this paper, they used the IRAF task
\verb+mkobject+, which, as we described in \S\ref{sec_sims}, is inexact for the
inner pixels of a simulated galaxy light profile. According to initial tests, the
differences in the profiles can lead to systematic errors of up to 10-20\% in the
fitting parameters using \galfit\, due to the lack of oversampling of the inner
pixels when using at least our settings of artdata parameters, and depending on the
exact profile parameters. As we used our own simulation script in IDL, improving the
profile from IRAF might be possible by using different parameters settings.

\item In their paper, Pignatelli et al. allow the sky value to be fitted
as a free parameter for all three codes. We argue in this paper that this is a
non-optimal way to run {\it any} galaxy fitting code: not only would one be subject
to errors from irregularities around a \sersic\ profile, but also the tests shown in
this paper show that fitting the sky level as an additional parameter leads to
significantly worse fits (especially in the case of \gimtwod). Estimating a value
for the sky {\it before} running the fitting codes and keeping this value fixed
returns more accurate galaxy parameter values.

\item Pignatelli et al. state that all automatic tools are likely to have
problems with blended objects. Like them, we find that deblending is necessary when
setting up fitting routines. Masking out blended objects, while better than doing
nothing at all, still leads to significantly biased results: this appears to lie at
the root of \gimtwod's difficulties in fitting some simulations (Fig.
\ref{fig_neighbor}). We find, furthermore, that if one fits multiple galaxies
simultaneously (as is recommended when using \galfit), \galfit\ returns stable
unbiased galaxy parameters, even in strongly-blended cases (and in cases with
intermixed $n=1$ and $n=4$ galaxies in which $n=4$ galaxies play an important role
and which was not tested in the paper by \citet{Pignatelli}; compare to \S\
\ref{sec_results_fake} in this work). Their argument that \galfit\ does not deal
well with blended galaxies is an artefact of the mode in which they chose to use
\galfit, in particular the lack of simultaneous fitting of neighboring objects.

\item Pignatelli et al. only show the \gimtwod\ results for $n$=4
galaxies; according to our tests these are the hardest galaxies to reliably fit, and
showing only those galaxies leads to a false impression of the frequency and
severity of \gimtwod's difficulties with nearby neighbors. Also, it seems that
\citet{Pignatelli} have used the standard setup for \gimtwod, which, according to
our tests, behaves poorly for faint galaxies: the influence of this decision on
their fitting results is unknown.
\end{itemize}

\section{GEMS GALFIT results} \label{sec_gems_results}
In this section, we present the \galfit\ F850LP-band fitting results of all 41,495
\gems\ objects that were found by \sex. We include fit results for all {\it unique}
objects, be they stars or galaxies. Some objects appear on two or more \gems\
frames; in this case the fit results for the images lying furthest from the frame
edge was taken. Table \ref{tab_fit_results} shows the 10 first objects in the
catalog and gives the format of the catalog published in the online version of this
paper. It includes the following values:
\begin{enumerate}
\item RA (1), Dec (2): RA and DEC, given by \sex (J2000).

\item tile (3): the \gems\ tile in which the galaxies appears

\item Snum (4): the \sex\ catalog number of this object

\item \gems ID (5): the identification of the galaxy within the \gems\
project.

\item PosX (6), PosY (7): the position [pixels] of the galaxy in this
given \gems\ tile.

\item sky (8): The background pedestal as returned by \galapagos\ and
used during the fit with \galfit.

\item The \galfit\ results (9-13): magnitude, halflight-radius $r_{e}$,
\sersic\ index $n$, Axis Ratio $b/a$, and position angle (both with respect to the
image, $PA_{im}$, and with respect to the WCS, defined north-to-east, $PA_{WCS}$) as
well as their `uncertainties'. These uncertainties are {\it not} the errorbars
returned by \galfit; as shown in \S\ref{sec_error_bars} these errorbars do not
reflect the true uncertainty of the fit. We use a statistical method to derive the
error estimates from our simulations. We first estimate from our simulations the
scatter of the distribution (of the $n=1$ and $n=4$ galaxy sample, respectively) at
the given surface brightness $\mu$ of the real object for $n=1$ and $n=4$
simulations. Then, we perform a linear interpolation between the $\sigma(n=1,
\mu=\mu_{obs})$ and $\sigma(n=4, \mu=\mu_{obs})$ to estimate $\sigma(n=n_{obs},
\mu=\mu_{obs})$. We do not extrapolate; galaxies with $n<1$ are given the value of
the $n=1$ sample, $n>4$ galaxies the value of the $n=4$ sample. We further adopt a
minimum uncertainty for each fitting parameter (0.01 mag for mag, 0.01 pixels for
$r_{e}$, 0.01 for $n$, 0.001 for $b/a$ and 0.1 deg for PA). In the table published
online and on the \gems\ webpage, the uncertainties are stored in extra columns.

\item $f_{con}$ (14): A flag showing which fits ran into any of the
fitting constraints (0: fit ran into constraint, 1: fit did not run into any of the
constraints).

\item $f_{sci}$ (15): A flag showing which galaxies would be selected
according to the selection criteria given in \citet{Barden2005} (0: object would not
be selected for analysis, 1: object would be selected for analysis). The primary
effect of the selection is to discard stars and very low surface brightness objects.
\end{enumerate}

\begin{figure*}[htb]
\begin{center}
\includegraphics[width=12cm,angle=0]{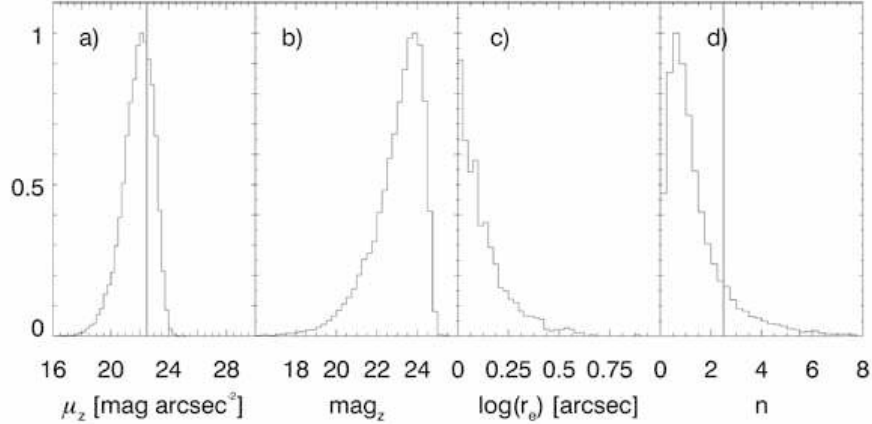}
\end{center}
\caption{This figure shows histograms for 23,187 objects from the \gems\ survey that
were fitted without running into constraints when fitted by \galfit\ and selected
according to the selection criteria in \citet{Barden2005}. From left to right we
show surface brightness $\mu$, apparent magnitudes, apparent sizes $r_{e}$
(logarithmic scale) and \sersic\ index $n$. For comparison we overplotted the
surface brightness of the sky as a vertical line in the leftmost plot and the cut of
$n$=2.5 in the rightmost plot, which is frequently used to distinguish between disk-
and bulge-dominated galaxies in an automated fashion.
\label{fig_gems_results_histos}}
\end{figure*}

As is clear from Fig.\ \ref{fig_gems_results_histos} and \ref{fig_completeness}, the
catalog has strongly varying completeness, primarily as a function of surface
brightness.  Many applications of the \gems\ catalogs require a good understanding
of these completeness properties.  In \citet{Barden2005} and \citet{McIntosh2005},
we used the simulations presented in this paper to quantify the effects of
completeness. Accordingly, we have made extensive suites of simulation catalogs
available to interested users on the \gems\ webpage to allow detailed examination of
systematic errors in fitting and sample completeness. These issues are discussed in
substantially more detail in \citet{Rix}, \citet{Barden2005} and
\citet{McIntosh2005}.

\begin{figure*}[htb]
\begin{center}
\includegraphics[width=8cm,angle=0]{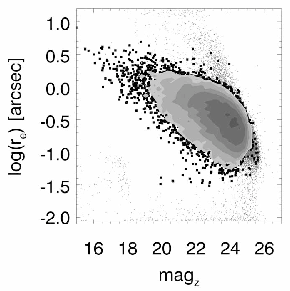}
\includegraphics[width=8cm,angle=0]{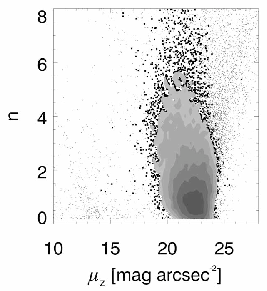}
\end{center}
\caption{Here we show parameter distributions for all galaxies in the catalog
published in Table \ref{tab_fit_results}, excluding the ones where the fit ran into
any of the fitting constraints. Galaxies plotted in grey do not end up in the
science sample according to the selection criteria used in \citet{Barden2005}.
Galaxies plotted in black and indicated by the contours pass this selection. In both
plots one can see that the galaxies thrown out are mostly faint low surface
brightness galaxies. \label{fig_gems_results_1}}
\end{figure*}

Figure \ref{fig_gems_results_1} shows the parameter distribution of the subset of
34,638 objects for which the fit did not run into fitting constraints. Galaxies
plotted in black and indicated by the contours would pass the selection in
\citet{Barden2005} . One can see that galaxies discarded (plotted in grey) are
mostly faint, low surface brightness galaxies. Another important class of objects
thrown out of the sample are objects with either very small sizes or relatively
small sizes at high magnitudes. These are identified as stars (or saturated stars)
by the automated selection criteria in \citet{Barden2005}. Although all these
objects are still included in Table \ref{tab_fit_results}, one should be very
careful when using their fitting results. All these galaxies are indicated by
$f_{sci}=0$.

Figure \ref{fig_gems_results_histos} shows histograms of the most important
parameters (surface brightness $\mu$, apparent magnitude, apparent size $r_{e}$ and
\sersic\ index $n$) for the subset of 23,187 objects that would be selected
according to the cuts given in \citet{Barden2005}.

\section{Conclusions} \label{sec_concl}
In this paper, we have tuned and tested two parametric galaxy fitting codes --
\galfit\ and \gimtwod\ -- for fitting single \sersic\ light profiles to both
simulated and real data. Our conclusions are the following:
\begin{itemize}

\item  The performance of both \galfit\ and \gimtwod\ is a strong
function of how the codes are set up; in particular, studies using
different setups of parametric fitting codes may arrive at different
conclusions about those codes if not properly or optimally used.

\item The recommended setup of \gimtwod, using \verb+`dobkg'+=`yes' and
\verb+`initparams+=`yes' is unable to recover the input parameter values of
simulated $n=4$ galaxies that were fainter than the sky surface brightness. We {\it
strongly} discourage users from using these settings, but to instead fix the
background to the value local of each galaxy as given by \sex\, and to input very
wide model parameter limits. This is very important if the \sex\ segmentation map
does not represent the true extent of a galaxy, as was the case for galaxies below
the sky surface brightness when using standard \sex\ configurations.

\item Both codes are able to fit (at least bright) $n=1$ galaxies reasonably well with
relatively little bias (Figure \ref{fig_disk_sims}). Concentrated $n=4$ galaxies are
substantially less straightforward to fit, owing to the large amount of light in the
faint outer parts of the galaxies. For bright $n=4$ galaxies, the behavior of
\galfit\ is better than that of \gimtwod\ (Figure \ref{fig_bulge_sims}); however,
parameters returned by \gimtwod\ are still not significantly biased. For galaxy
populations and clustering typical of medium-depth cosmological \HST\ surveys, there
are no large differences between results obtained using \gimtwod\ and \galfit\ for
these bright galaxies. For fainter galaxies, the performance of \galfit\ is
substantially better than that of \gimtwod. In the set of realistically mixed
simulations (Figure \ref{fig_gems_fake_disk} and \ref{fig_gems_fake_bulge}) of $n=1$
and $n=4$ galaxies, representing re-simulated \gems\ tiles, one can see
significantly different behavior of the two codes, especially in the recovery of the
\sersic\ index. \gimtwod\ results are systematically biased to higher \sersic\
indices, which, in automated galaxy classification using the \sersic\ index to
distinguish early- from late-type galaxies, would lead to systematic
misclassification of a subsample of (faint) galaxies.

\item The errorbars given by both codes underestimate the true
uncertainty of the fit by a large factor. One has to use a different approach to
derive more realistic errorbars.

\item Our testing demonstrated that how a code treats neighboring
galaxies can be of great importance. \gimtwod\ only masks out neighbors, which in
the tests we ran could lead to poor fitting results for strongly blended objects.
\galfit, in contrast, is able to simultaneously fit many objects, and when used in
that mode seems to be relatively robust to contamination by neighbors. For this
reason, we caution users interested in strongly clustered galaxies against using
\gimtwod\ without extensive prior testing.

\item Both \galfit\ and \gimtwod\ are self-consistent and show no
discernable dependence on image depth when comparing fitting results from \gems\ and
\goods\ data.

\item Our tests on deep and shallow data show that real galaxies are indeed reasonably
well described by general \sersic\ light profiles.

\item \galfit\ works best using an isophotal sky value given by
\galapagos. If this is not possible, using \galfit\ to internally derive a sky value
is significantly better than fixing the sky to a local value given by \sex.

\item On the balance, we would tend to recommend \galfit\ for single
\sersic\ profile fitting in medium-depth \HST/\ACS\ data, as \galfit\ results are
not only somewhat more reliable in the mean, but also have lower scatter and less
sensitivity to contamination by neighbors than \gimtwod.
\end{itemize}

\acknowledgments We thank the anonymous referee for his useful comments and his kind
report. E.F.B. was supported by the Deutsche Forschungsgemeinschaft's Emmy Noether
Programme. D.H.M. acknowledges support from the National Aeronautics and Space
Administration (NASA) under LTSA Grant NAG5-13102 issued through the Office of Space
Science. S. Koposov was supported by SFB 439 (Collaborative research center 439
"Galaxies in the young universe"). Marco Barden was supported by the
"Bundesministerium f\"ur Bildung und Forschung" through DLR ("Deutsches Zentrum
f\"ur Luft und Raumfahrt") under grant 50 OR 0401.

\clearpage \thispagestyle{empty}
\begin{landscape}
\thispagestyle{empty}
\begin{deluxetable*}{lllcrrrrrrrrrrr}
\tablewidth{0pt} \tablenum{9} \tabletypesize{\scriptsize} \tabletypesize{\tiny}
\tablecaption{\galfit\ fitting results for all \gems\ galaxies}
\tablehead{\colhead{RA} & \colhead{DEC} & \colhead{tile} & \colhead{Snum} &
\colhead{\gems\ ID} & \colhead{PosX} & \colhead{PosY} & \colhead{sky} &
\multicolumn{5}{c}{\galfit\ fitting results} & \colhead{$f_{con}$} &
\colhead{$f_{sci}$}\\
\cline{9-13}\\
\colhead{} & \colhead{} & \colhead{} & \colhead{} & \colhead{} & \colhead{} &
\colhead{} & \colhead{} & \colhead{mag} & \colhead{$r_{e}$}
& \colhead{$n$} & \colhead{$b/a$} & \colhead{$PA$ image/WCS} & \colhead{} & \colhead{}\\
\colhead{[deg]} & \colhead{[deg]} & \colhead{} & \colhead{} & \colhead{} &
\colhead{[pix]} & \colhead{[pix]} & \colhead{[cnt]} & \colhead{[mag]} &
\colhead{[pix]} & \colhead{} & \colhead{} & \colhead{[deg]} & \colhead{} & \colhead{}\\
\colhead{(1)} & \colhead{(2)} & \colhead{(3)} & \colhead{(4)} & \colhead{(5)} &
\colhead{(6)} & \colhead{(7)} & \colhead{(8)} & \colhead{(9)} & \colhead{(10)} &
\colhead{(11)} & \colhead{(12)} & \colhead{(13)} & \colhead{(14)} & \colhead{(15)}}
 \startdata
 53.316325 & -28.059058 & s9z01A &  1 & GEMSJ033315.92-280332.6 & 5752.64 & 721.57 & 18.120 & 13.91$\pm$0.01 &  0.32$\pm$0.01 & 6.63$\pm$0.01 &  0.777$\pm$0.011 &  3.8(  1.6)$\pm$0.1 & 1 & 0\\
 53.334888 & -28.062040 & s9z01A &  2 & GEMSJ033320.37-280343.3 & 3786.09 & 368.55 & 18.120 & 15.57$\pm$0.01 &  0.30$\pm$0.02 & 3.39$\pm$0.01 &  0.938$\pm$0.001 &  84.0( 81.8)$\pm$0.1 & 0 & 0\\
 53.340850 & -28.062310 & s9z01A &  5 & GEMSJ033321.80-280344.3 & 3154.69 & 337.72 & 18.056 & 21.58$\pm$0.09 & 13.93$\pm$0.11 & 0.37$\pm$0.26 &  0.745$\pm$0.056 &  21.4( 19.2)$\pm$4.4 & 1 & 1\\
 53.366943 & -28.062952 & s9z01A &  6 & GEMSJ033328.07-280346.6 &  391.45 & 266.85 & 18.355 & 24.14$\pm$0.10 &  5.21$\pm$0.12 & 0.97$\pm$0.27 &  0.719$\pm$0.059 & -47.8(-50.0)$\pm$4.6 & 1 & 1\\
 53.340514 & -28.061939 & s9z01A &  9 & GEMSJ033321.72-280343.0 & 3190.41 & 382.16 & 18.149 & 21.13$\pm$0.32 & 67.08$\pm$0.35 & 5.36$\pm$1.02 &  0.558$\pm$0.054 & -60.0(-62.2)$\pm$5.8 & 1 & 0\\
 53.370589 & -28.062339 & s9z01A & 10 & GEMSJ033328.94-280344.4 &    5.61 & 341.22 & 18.426 & 23.22$\pm$0.12 & 19.79$\pm$0.14 & 0.66$\pm$0.29 &  0.302$\pm$0.064 & -77.1(-79.3)$\pm$5.1 & 1 & 0\\
 53.342879 & -28.061422 & s9z01A & 11 & GEMSJ033322.29-280341.1 & 2940.04 & 444.77 & 18.098 & 23.05$\pm$0.11 & 14.03$\pm$0.13 & 1.00$\pm$0.28 &  0.619$\pm$0.063 & -38.8(-41.1)$\pm$5.1 & 1 & 1\\
 53.342659 & -28.060898 & s9z01A & 12 & GEMSJ033322.24-280339.2 & 2963.58 & 507.56 & 18.171 & 20.05$\pm$0.01 &  0.30$\pm$0.17 & 1.17$\pm$0.92 &  0.140$\pm$0.001 &  52.4( 50.1)$\pm$0.4 & 1 & 0\\
 53.347315 & -28.061235 & s9z01A & 13 & GEMSJ033323.36-280340.4 & 2470.43 & 468.25 & 18.223 & 22.96$\pm$0.08 &  6.30$\pm$0.10 & 1.18$\pm$0.27 &  0.804$\pm$0.051 &  57.5( 55.2)$\pm$4.1 & 1 & 1\\
 53.341322 & -28.059825 & s9z01A & 14 & GEMSJ033321.92-280335.4 & 3105.38 & 636.05 & 18.144 & 21.56$\pm$0.10 & 43.40$\pm$0.12 & 0.45$\pm$0.27 &  0.123$\pm$0.059 &  50.7( 48.4)$\pm$4.7 & 1 & 1\\
\enddata \tablecomments{This table shows the first 10 objects of the \gems\
fitting results published in the online version of this paper. For every object that
was found by \sex\ we give RA, DEC, the \gems\ tile name, the \sex\ number of that
object, the \gems ID (containing RA and DEC) as well as the X and Y position on the
\gems\ tile, so that identification of objects is easily possible. Furthermore we
give the isophotal sky value that was used during the fit and the fitting results of
the 5 key parameters mag (in apparent F850LP-band magnitudes), $r_{e}$ (in pixels of
0.03"), \sersic\ index $n$, axis ratio $b/a$  and position angle PA in respect to
the image (counted counterclockwise from vertical line according to the convention
within \galfit) and the world coordinate system (in brackets, north through east).
Uncertainty estimates are derived from the simulations, following \S
\ref{sec_error_bars}. The last two columns show two different flags showing which
fits ran into fitting constraints ($f_{con}=0$) and which ones would make it into
the final galaxy sample used for science according to the selection criteria given
in \citet{Barden2005} ($f_{sci}=1$).\\
The complete version of this table is in the electronic edition of the Journal. The
printed edition contains only a sample. \label{tab_fit_results}}
\end{deluxetable*}
\clearpage
\end{landscape}

\end{document}